\documentclass[traditabstract]{aa}
\usepackage{graphicx}                    
\usepackage{bm}
\usepackage{times}                    
\usepackage{natbib}                     
 \bibpunct{(}{)}{;}{a}{}{,} 
\usepackage{amssymb}                    
\usepackage{color}                       
\usepackage{url}                         
\usepackage{amsmath} 

\def\beg{\begin{eqnarray}}
\def\ende{\end{eqnarray}}
\def\lsim{\lower.4ex\hbox{$\;\buildrel <\over{\scriptstyle\sim}\;$}}
\def\gsim{\lower.4ex\hbox{$\;\buildrel >\over{\scriptstyle\sim}\;$}}

\newcommand{\Prr}{\mbox {Pr}}

\renewcommand{\vec}[1]{\mbox{\boldmath $#1$}}
 
\def \Qrt{Q_{r\theta}}
\def \Qrf{Q_{r\phi}}
\def \Qtf{Q_{\theta\phi}}
\def \Qxy{Q_{xy}}

\def \Qyz{Q_{yz}}

\def \dd{{\textrm d}}

\def \Om  {{\it \Omega}}
\def \Omst{{\it \Omega}}

\def \nuT{\nu_ \Vert}
\def \hatnuT{{\nu_\bot}}

\def\gsim{\lower.4ex\hbox{$\;\buildrel >\over{\scriptstyle\sim}\;$}} 
\def\lsim{\lower.4ex\hbox{$\;\buildrel <\over{\scriptstyle\sim}\;$}} 

\renewcommand{\vec}[1]{\mbox{\boldmath $#1$}}

\def\aap{Astronomy \& Astrophysics}

\def\apjs{Astrophys. J. Suppl.}

\begin{document}



\title{Antisolar differential rotation of slowly rotating cool stars}
\titlerunning{Antisolar differential rotation of  stars}
%
\author{G.~R\"udiger\inst{1,2} \and M.~K\"uker\inst{1} \and P. J. K\"apyl\"a\inst{3,4} \and K.G. Strassmeier\inst{1,2}}


%
  \institute{Leibniz-Institut f\"ur Astrophysik Potsdam (AIP), An der Sternwarte 16, D-14482 Potsdam, Germany,
    email: gruediger@aip.de
 \and
  University of Potsdam, Institute of Physics and Astronomy, Karl-Liebknecht-Str. 24-25, 14476 Potsdam, Germany
 \and
    Georg-August-Universit\"at G\"ottingen, Institut f\"ur Astrophysik, Friedrich-Hund-Platz 1, D-37077 G\"ottingen, Germany
  \and
    ReSoLVE Centre of Excellence, Department of Computer Science, Aalto University, PO Box 15400, FI-00076 Aalto, Finland}

\date{Received; accepted}

\abstract{Rotating stellar convection transports angular momentum towards the equator, generating the characteristic equatorial acceleration of the solar rotation while 
the radial flux of angular momentum is always inwards. 
New numerical box simulations for the meridional cross-correlation  $\langle u_\theta u_\phi\rangle $, however,   reveal   the angular momentum transport towards the poles for slow rotation and towards the equator for fast rotation.  The explanation is that for slow rotation  a  negative radial gradient of the angular velocity always appears, which in combination with  a so-far neglected rotation-induced off-diagonal eddy viscosity term $\hatnuT$ provides  ``antisolar rotation''  laws  with a decelerated equator.  Similarly, the simulations  provided  positive values for the  rotation-induced correlation  $\langle u_r u_\theta\rangle $,
which is relevant for the resulting  latitudinal temperature profiles (cool or warm poles)  for slow rotation and negative values  for fast rotation. 
 Observations  of the differential rotation of  slowly rotating stars  will therefore  lead to a   better understanding of the  actual stress-strain relation, the heat transport,  and the underlying rotating convection.
}


\maketitle
%
\section{Introduction} \label{Section1}
Fast stellar rotation together with turbulent convection leave an imprint on the stellar surface in the form of starspots
\citep{S03,S09}.
Nevertheless, numerous puzzles remain for the quantitative description of the concerted action of stellar rotation and magnetic-field amplification in cool late-type stars. Surface differential rotation in its solar and antisolar form is one of them. Direct numerical simulations and mean-field models deal with the impact of Reynolds stresses and thermal energy flows on angular momentum transport in rotating convection,  which are thought to be responsible for the observed meridional flows and differential rotation \citep[e.g.,][]{KR99,KM11,WK13,GY14, FM15}. \cite{KO11} showed that the meridional flow is distributed over the entire convection zone in slow rotators but retreats to the convection zone boundaries in rapid rotators. Mean-field models had already predicted antisolar differential rotation for stars with fast meridional circulation \citep{KR04}. Tracking sunspot and starspot migration from spatially resolved solar and stellar disk measurements \citep[e.g.,][]{KC15} provided us with direct observations of differential rotation also in its antisolar form (polar regions rotate faster than the equatorial regions) meaning that models can now be tested against observations.

Differential rotation from spot tracking is no longer confined to solar observations \citep[for a summary of solar differential rotation tracing measurements see, e.g.,][]{WB10}. Tracking starspot migration from spatially resolved stellar disk measurements (Doppler imaging) is a method to directly study stellar surface differential rotation. After pioneering work on image cross-correlation and smeared Doppler imaging by \cite{DC97},
\cite{BC05} concluded that differential rotation decreases
with effective temperature and rotation. We have now a number of (active) stars where differential rotation has been detected directly by means of Doppler imaging. A differential rotation versus rotation-period relationship from Doppler-imaging results was suggested by \cite{KO17} in the form $\delta \Om\simeq \pi/100$~rad/day, where $\delta \Om$ is the equator-pole difference of the angular velocity. Such a  very weak dependence of $\delta\Om$ on the stellar rotation rate of dwarf stars and giants  was first predicted by \cite{KR99}.

Space-based ultra-high-precision time-series photometry allowed confirmation of the temperature dependency of surface differential rotation for stars over a wide range in the Hertzsprung-Russel  diagram \citep{RR13}. The observed $\delta\Om$ only varied from 0.079 rad/day for cool stars ($T_{\rm eff} = 3500$~K) to 0.096 rad/day for $T_{\rm eff} = 6000$~K, which is a rather weak variation of the observed differential rotation on the rotation periods. Further, the hotter stars show stronger differential rotation, peaking at the F stars where there is still a significant convective envelope but only comparably weak magnetic activity. On the contrary,   despite their  small differential rotation  M stars appear to be highly dynamo-efficient \citep{GM13}.
  The overall rate of rotation plays an observationally biasing role here because smaller stars rotate much faster than bigger ones. Comparison of the large set of differential rotation measurements from the {\sc Kepler} mission with the theoretical predictions by the $\Lambda$ effect theory  showed a fair agreement and gives us confidence in applying the method to a particular star. However, the photometric data do not contain information on the sign of differential rotation (but see  \cite{RA15} for possible exceptions) and further spectroscopic time-series data are needed.

\cite{BB18} report the asteroseismic detection  of surface rotation laws of solar-type stars with rather large equator-pole differences of the angular velocity. Among the sample of  40 stars there are up to 10 candidates for antisolar differential rotation with a weak anticorrelation to rapid rotation. Asteroseismology  for the two solar analogs 16 Cyg A and B (which rotate slightly faster than  the Sun) provided positive equator-pole $\Om$~differences  only slightly larger than the solar value \citep{BB19}. We are therefore tempted to study the possibility of antisolar rotation mainly for  slowly rotating main sequence stars. Also, the results of numerical simulations     by \cite{G77} and  later by \cite{GY14,BP09,GS13,GY14,KK14}   suggest that stars with slow rotation possess antisolar rotation laws. Most recently, \cite{VW18} reported 3D simulations of turbulent convection for rotation rates of the solar value and faster. A transition of antisolar to solar-like differential rotation happened for about 1.8 solar rotation rates (their Fig. 5). Simultaneously, the geometry of the dynamo-excited large-scale magnetic field became nonaxisymmetric. Even more important for the understanding of the differential rotation problem is that the radial rotation shear also simultaneously changed from negative (``subrotation'') to positive (``superrotation''). We demonstrate in the present paper that indeed the subrotation $\Om$-profile generically belongs to the antisolar-rotation phenomenon of decelerated equators.


We  present numerical 3D box-simulations of outer stellar   convection zones subject to slow rotation with a fixed Prandtl number. We then check if the resulting azimuthal cross-correlations generate solar-type or antisolar-type rotation laws. Our analytical  differential rotation model is described in Sect. \ref{Differential} while the simulations of the rotating convection boxes are presented in Sect.~\ref{Rotating}. The  results in terms of the eddy viscosity tensor for the angular momentum flow are   given in Sect.~\ref{Eddy} and they are finally discussed in Sect.~\ref{Discussion}.
%
\section{Differential rotation}\label{Differential}
%
{In the following our theory of differential rotation in shellular  convection zones  in the mean-field  hydrodynamic approach is briefly reviewed. This is mainly the theory of 
angular momentum conservation
including meridional flow and  Reynolds stress, that is,
\begin{equation}
\frac{\partial}{\partial t} (\rho r^2\sin^2\theta \Om) + \nabla\cdot
\left\{\rho  r^2\sin^2\theta \Om {\vec U} + \rho  r\sin\theta
\langle  u_\phi \vec{u}\rangle\right\} = 0,
\label{angtrans}
\end{equation}
where $\rho$ is the mass density, $\Om$  the angular velocity,
$ {\vec{U}}$ is the ensemble average of the fluid velocity,
and $\vec{u}$ are the fluctuating parts of the flow. Equation (\ref{angtrans}) describes the contributions of the two main transporters of angular momentum in (unmagnetized) rotating convection zones. The cross-correlations $Q_{r\phi}=\langle u_r(\vec{x},t) u_\phi\rangle(\vec{x},t) $ and $Q_{\theta\phi}=\langle u_\theta(\vec{x},t) u_\phi\rangle(\vec{x},t) $  describe the radial and latitudinal turbulent transport of angular  momentum \citep{R89}. In  the simplest case they can be parametrized via the diffusion approximation,
\beg
\begin{split}
Q_{r\phi}=-\nuT& \sin\theta \frac{r\partial \Om}{\partial r}, \ \ \ \ \ \ \ \ Q_{\theta\phi}=-\nuT \sin\theta \frac{\partial \Om}{\partial \theta},
\end{split}
\label{Bou}
\ende
with $\nuT$ being the positive eddy viscosity\footnote{$\nuT=\nu_1\equiv \nu_{\rm T}$ in the notation of \cite{KP94}}. In this approximation the two cross-correlations 
would vanish for uniform rotation. If, therefore, under certain circumstances the 
cross-correlations for uniform rotation do not vanish, the Boussinesq formulation (\ref{Bou}) can no longer be true and uniform rotation cannot form a solution of Eq. (\ref{angtrans}) in rotating turbulence fields.  Indeed theory, simulation, and observation suggest that  large-scale  stellar convection produces finite values for the two mentioned 
cross-correlations, this phenomenon being referred to as the ``$\Lambda$ effect''. For  the  Sun as a rapid rotator (compared with the typical correlation times)  \cite{H13} indeed reported positive  latitudinal cross-correlations for the northern hemisphere and   negative  latitudinal cross-correlations for the southern  hemisphere in contradiction to the simple diffusion approximation  (\ref{Bou}) which would provide  opposite signs.}

The symmetry properties of the cross-correlations $ Q_{r\phi}$ and $ Q_{\theta\phi}$ differ from
those  of all other components of the one-point correlation  tensor, 
\beg 
Q_{ij}=\langle u_i(\vec{x},t) u_j(\vec{x},t) \rangle,
\label{Q}
\ende 
of a rotating turbulence field. 
While $Q_{r\phi}$ and $ Q_{\theta\phi}$ are antisymmetric with respect to the transformation $\Om\to -\Om$, all other correlations are not. The turbulent angular momentum transport is thus odd in $\vec{\Om}$ while the other two tensor components -- the cross-correlation $Q_{r \theta}$ included -- are even in $\vec{\Om}$. It is easy to show that $Q_{r\phi}$ is symmetric with respect to the equator if the averaged flow  is   also symmetric. In this case, the component $Q_{\theta\phi}$ is antisymmetric with respect to the equator.
 These rules can be violated if, for example, a magnetic field exists whose amplitudes are different  in the two hemispheres.

One can also  show that isotropic turbulence even under the influence of  rotation does not lead to finite values of  $Q_{r\phi}$ and $ Q_{\theta\phi}$. With a preferred (radial) direction $\vec g$,  a tensor $(\epsilon_{ikl} g_j+ \epsilon_{jkl}g_i)g_k\Om_l$ linear in $\vec{\Om}$ can be formed that has nonvanishing  $r \phi$ and $\theta \phi$ components.
Rotating anisotropic turbulence is therefore able to transport
angular momentum.
The spherical coordinates $(r, \theta,\phi)$ are used in this paper  if the global system is concerned  while $(x,y,z)$ represent these  coordinates in a Cartesian box geometry.

  For the zonal fluxes of angular momentum we write
\beg
\begin{split}
Q_{r\phi}=-\nuT& \sin\theta \frac{r\partial \Om}{\partial r}+\\\ & +{\hatnuT}\Om^2\sin^2\theta \cos\theta \frac{\partial \Om}{\partial \theta}
+\nuT V\sin\theta \Om
\end{split}
\label{dr01}
\ende
for the radial flux and
\beg
\begin{split}
Q_{\theta\phi}=-\nuT& \sin\theta \frac{\partial \Om}{\partial \theta}+\\\ & +{\hatnuT}\Om^2\sin^2\theta\cos\theta \frac{r\partial \Om}{\partial r}+\nuT H \cos\theta \Om
\end{split}
\label{dr02}
\ende
for the meridional flux.
Here the first terms  come from the  Boussinesq    diffusion approximation with $\nuT$  as the   eddy viscosity while $V$ and $H$ form the components of the $\Lambda$ tensor describing the angular momentum transport of rigidly rotating anisotropic turbulence. The terms with  ${\hatnuT}$ follow from the fact that a viscosity tensor connects the Reynolds stress with the deformation tensor\footnote{$\hatnuT= \nu_2$ in the notation of \cite{KP94}}.   
The nondiffusive terms in the zonal fluxes (\ref{dr01}) and (\ref{dr02})  can be written by means of the stress-strain tensor relation $Q_{i\phi}=-{\cal N}_{ij}\nabla_j
\Om$ with
\beg
{\cal N}= r
 \left( \begin{array}{lll}
\sin\theta\ \nuT & -\cos\theta\sin^2\theta \Om^2 {\hatnuT} & 0\\
-\cos\theta\sin^2\theta \Om^2\ \hatnuT   & \sin\theta\ \nuT &  0\\
0 & 0 & 0
\end{array}\right)
\label{nutensor}
\ende
\citep{R89}. The positive standard eddy viscosity $\nuT$  is quenched by fast rotation. On the other hand, \cite{KP94} showed  that for isotropic and homogeneous turbulence  $\hatnuT$ is positive. We use ${\hatnuT}$ to refer to the rotation-induced off-diagonal viscosity term. The off-diagonal viscosity
does not contribute at the poles or at  the equator.
We note that  this term  in Eq. (\ref{dr02})  transforms a positive (negative)  radial $\Om$ gradient into positive (negative) cross-correlations which -- as the solution of the equation for the  angular momentum conservation -- finally leads to  accelerated (decelerated) equators.
Hence,   the rotation law of   a  convection zone can never be only radius-dependent. After the Taylor-Proudman theorem the isolines of the angular velocity $\Om$ tend to become cylindrical so that (say) slower  rotation in the depth of the convection zone is transformed to polar deceleration (solar-type rotation). If, on the other hand,  the inner parts rotate faster than the outer parts  then automatically  the polar regions rotate  faster than the more equatorial regions (``antisolar rotation'')\footnote{In the linear-in-$\Om$ approximation  by  \cite{K63} a very similar mechanism is realized  via  meridional flow.}.

The expansion
\beg
V=\sum_{l=0}{ V^{(l)}\sin^{2l}\theta} \sin\theta, \ \ \ \ \ \ \ \ \  
H=\sum_{l=1}H^{(l)}\sin^{2l}\theta \cos\theta
\label{dr11}
\ende   
 is used for the normalized $\Lambda$ effect as in earlier papers. The  coefficients $V^{(l)}$ and $H^{(l)}$  describe the latitudinal profile of the $\Lambda$ effect. Quasilinear theory of rapidly rotating anisotropic turbulent convection  in the high-viscosity limit leads to  $-V^{(0)}=V^{(1)}=H^{(1)}>0 $ (with $V^{(l)}=H^{(l)} =0 $ for $l>1$) which implies  that 
 $Q_{r\phi}$ vanishes at the equator. For rigid rotation and in cylindric coordinates    $-H \sin\theta \cos\theta$ is the angular momentum flow in axial direction while the radial flux of angular momentum vanishes. The function $H=H(\Om)$ is positive definite, meaning that  the  angular momentum is exclusively transported from the poles to equator parallel to the rotation axis. $V^{(0)}$ is always negative \citep{RE05}. 
 \begin{figure}[htb]
 \centering
\includegraphics[width=2.75cm]{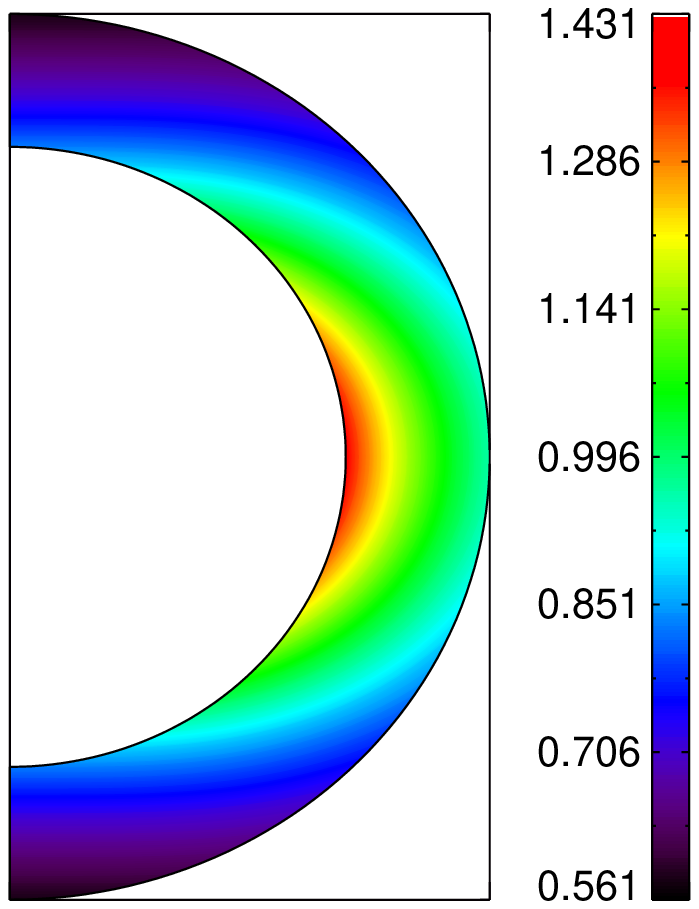}
\includegraphics[width=2.75cm]{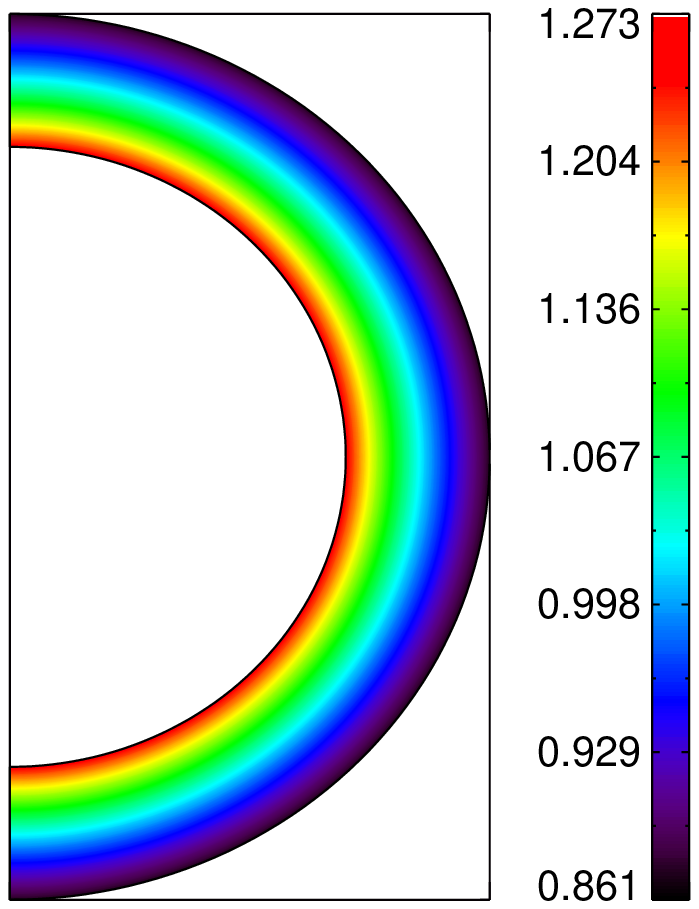}
\includegraphics[width=2.75cm]{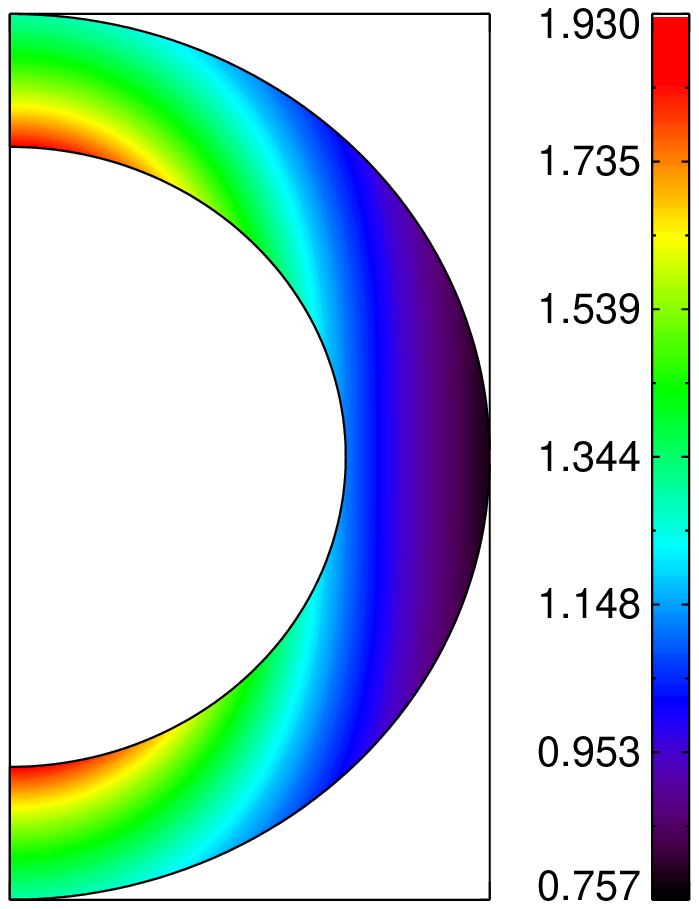}
\caption{Color-coded contours of the rotation law due to the $\Lambda$ effect with 
 $V^{(0)}=-1$ , $V^{(1)}=0$
 and $H^{(1)}=1$ (left),  $H^{(1)}=0$ (middle), $H^{(1)}=-1$ (right). Meridional circulation and the off-diagonal eddy viscosity  are artificially suppressed. Positive $H^{(1)}$ values lead to solar-type equatorial acceleration 
while  negative $H^{(1)}$ values lead to antisolar rotation profiles.  $\hatnuT=0$.}
\label{fig1}
\end{figure} 
 
 Either  of the above-mentioned theories and simulations of the radial $\Lambda$ effect lead to results of the form  $V\propto - \cos^2\theta$; that is, $V^{(0)}<0$ and $V=0$  at the equator. For slow rotation,  $V^{(l)}$ and $H^{(l)}$ with $l>0 $ become so small that a radial rotation law with 
 \beg
 \frac{{\rm d}\log\Om}{{\rm d}\log r}= V^{(0)}
 \label{dr12}
 \ende   
results. Negative $V^{(0)}$  values generally  lead to radial $\Om$ profiles with negative shear. In this case a meridional circulation is driven by the centrifugal force which at the surface transports angular momentum towards the poles (``counterclockwise flow''). Hence, the equator rotates slower than the mid-latitudes which automatically leads to  an antisolar rotation law with $\cos\theta \partial \Om/\partial\theta<0$ at the surface.
 
If neglecting meridional flow and $\hatnuT$, the Reynolds stress (\ref{dr11}) maintains a latitude-dependent surface rotation law $\Om=\Om(\theta)$
 with 
   \beg
 \frac{\delta\Om}{\Om}= -\frac{1}{2}\sum_{l=1}{\left( { d}\  V^{(l)} +  \frac{H^{(l)}}{l} \right)}
 \label{dr13}
 \ende   
  for the pole-equator difference of $\Om$ and with  the normalized thickness $ d$ of the convectively unstable  layer with   stress-free boundary conditions. Under the assumption that $V^{(l)}=H^{(l)}=0$ for  $l>1,$ antisolar rotation would only be  possible for   (formally) negative  $H^{(1)}$. 
  
  Figure \ref{fig1}  illustrates the consequences of  (\ref{dr13}). The equation of angular momentum is solved for  a negative $V^{(0)}$. For this demonstration, meridional circulation and the off-diagonal viscosity $\hatnuT$ have been  neglected.
The latitudinal $\Lambda$ effect represented  by $H^{(1)}$ is varied from 1 to $-1$. 
Not surprisingly, a negative pole-equator difference of the surface rotation law (solar-type differential rotation)  originates from  $H^{(1)}= 1$ (left panel). {For $H^{(1)}=0,$ a shellular rotation profile results.
Moreover, for  $H^{(1)}=-1$ the solar-type surface rotation law changes to an antisolar-type surface rotation law with positive pole-equator difference (right panel). At the same time the isolines of the angular velocity of rotation change from disk-like (left panel) to cylinder-like (right panel). This phenomenon is due to Reynolds stress  rather than  to Taylor-Proudman theorem. In the left plot the angular momentum  is transported by the $\Lambda$ effect along cylindrical planes to the equator causing the $\Om$-isolines to become disk-like. In the  middle plot the transport is radial, meaning that the $\Om$-isolines become shellular and in the right plot the angular momentum is  transported toward    the rotation  axis generating cylindrical $\Om$-isolines. 
 }
   \begin{figure}[htb]
 \centering
\vbox{
\hbox{ \includegraphics[width=2.99cm]{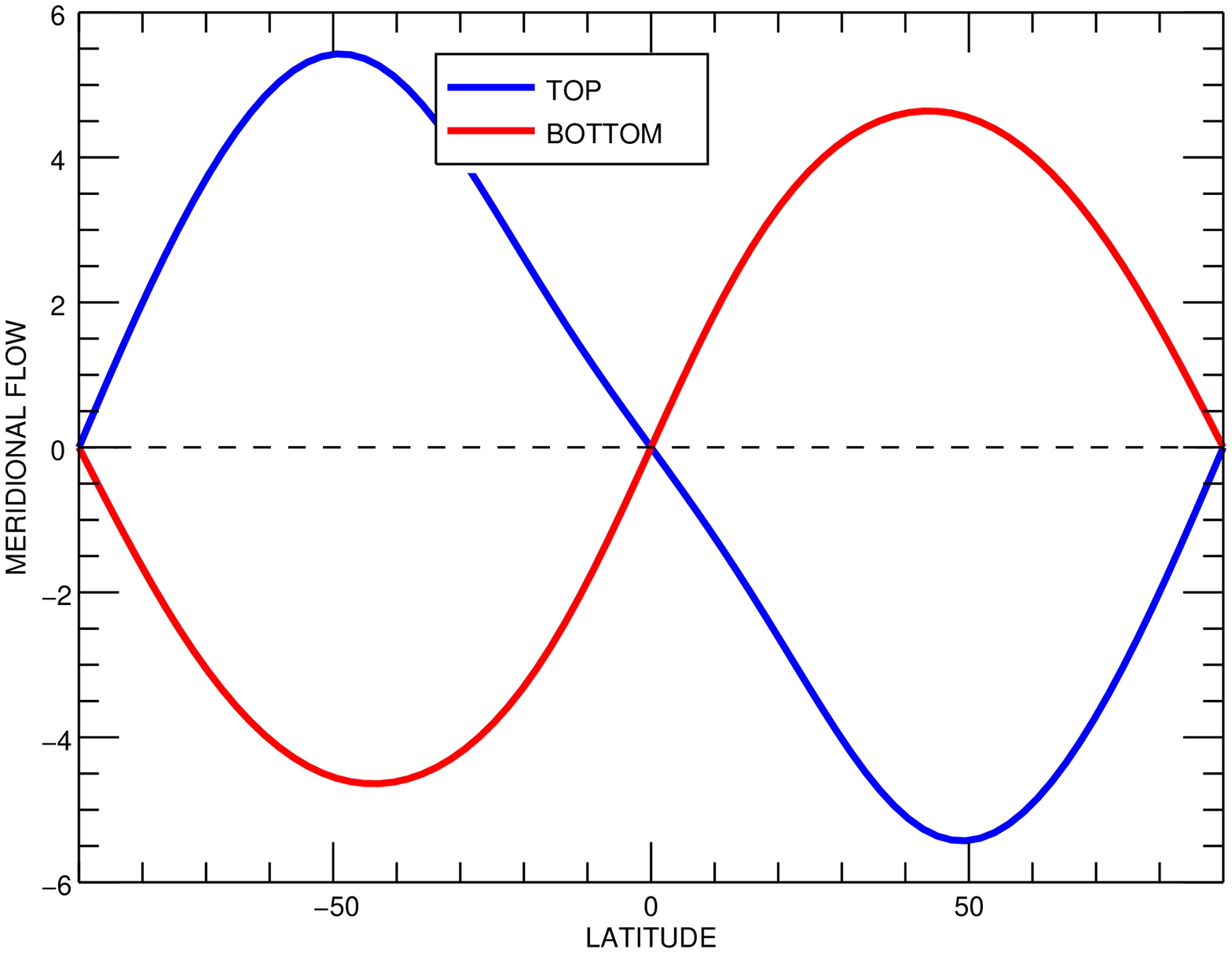}
\includegraphics[width=2.99cm]{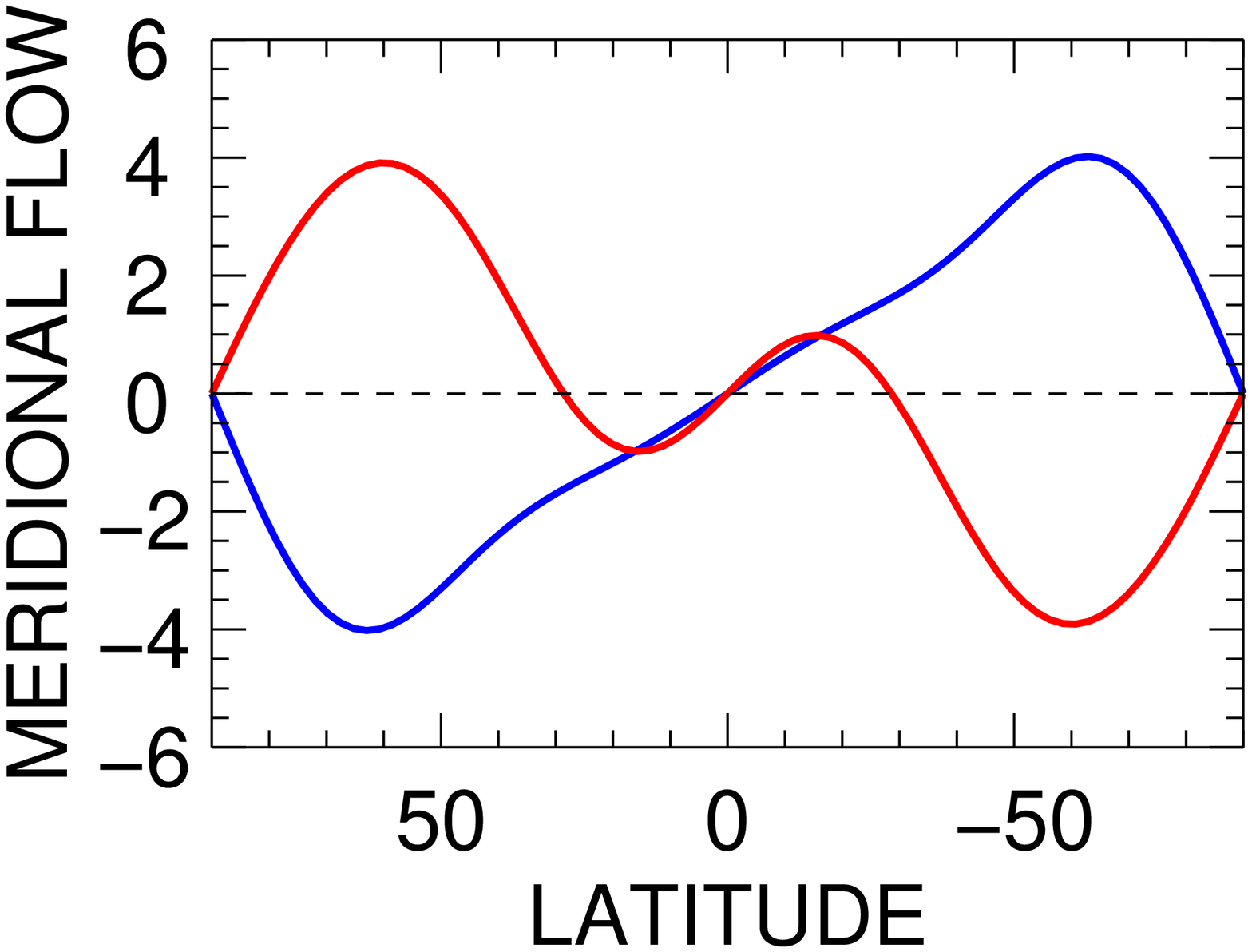}
\includegraphics[width=2.99cm]{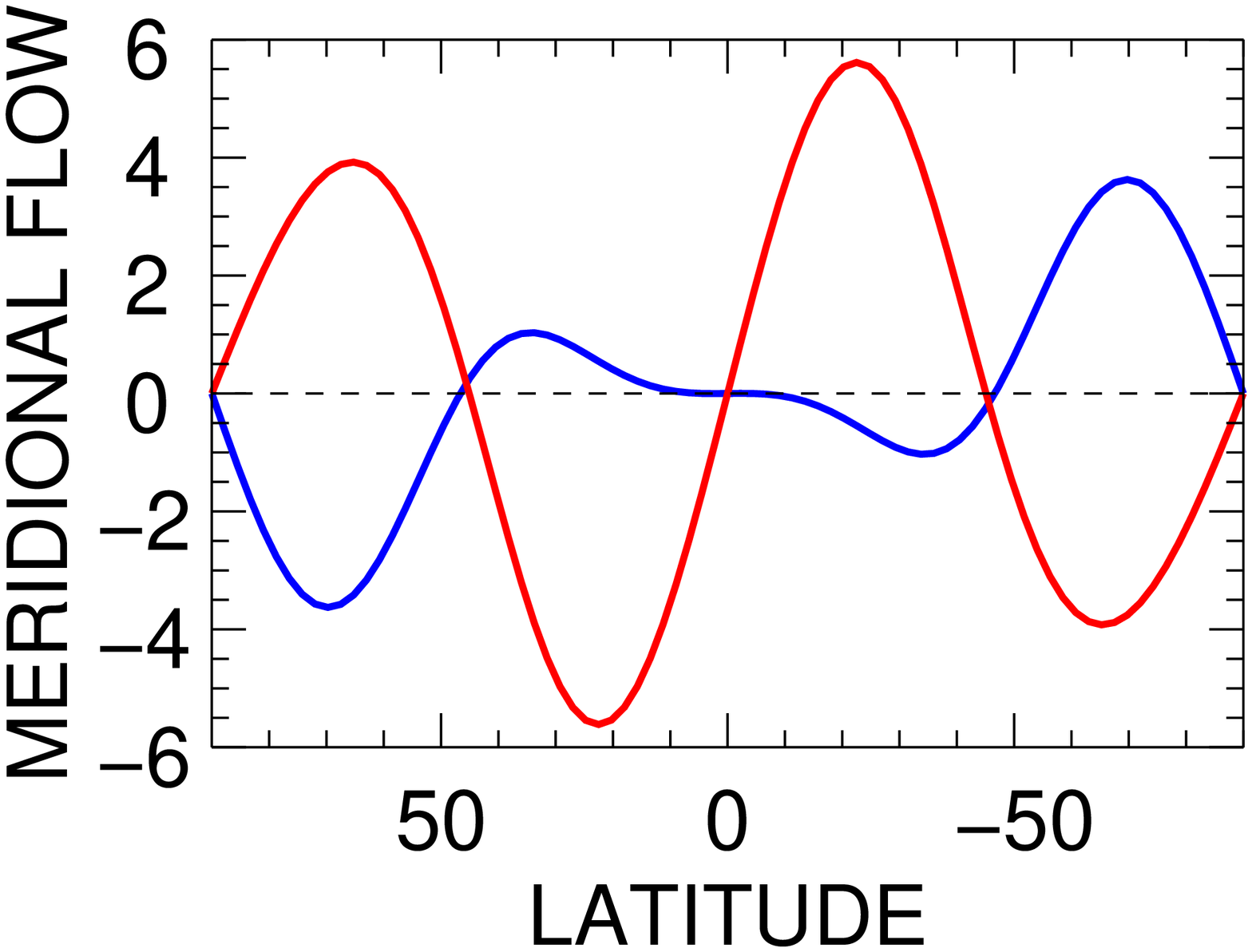}
}
\hbox{
 \includegraphics[width=2.9cm]{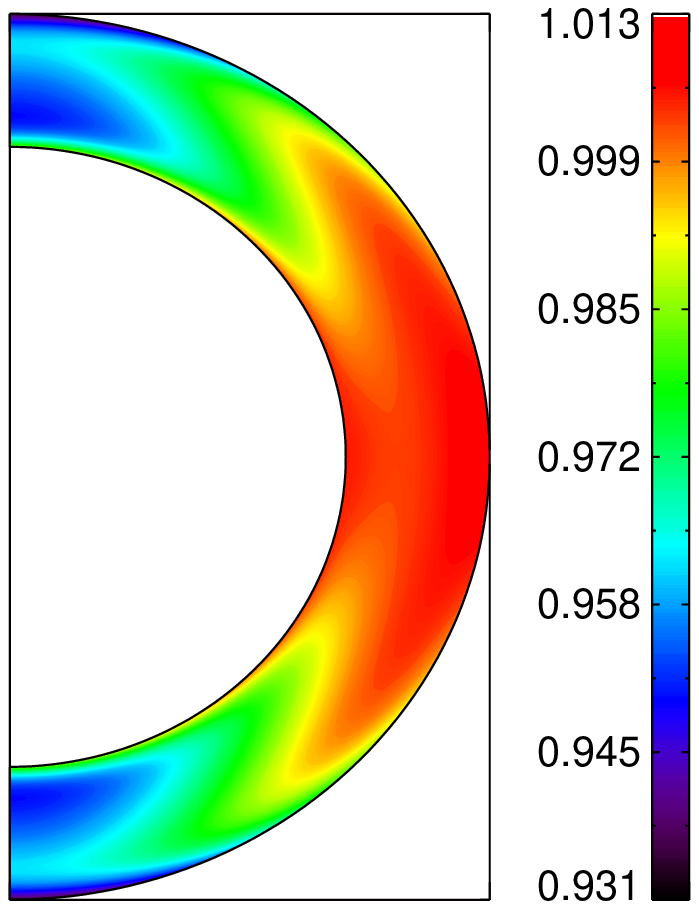}
\includegraphics[width=2.9cm]{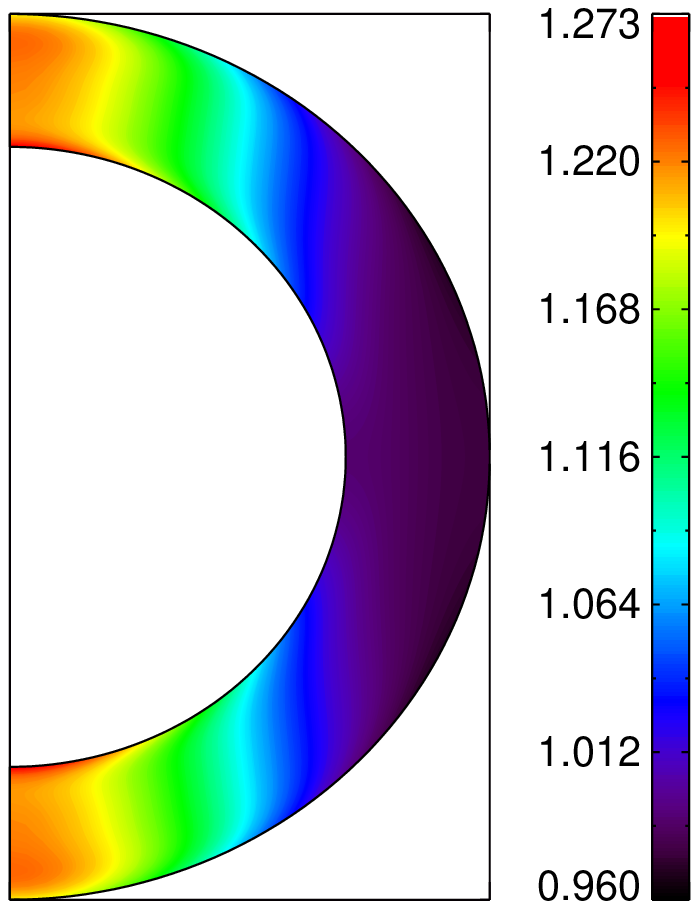}
\includegraphics[width=2.9cm]{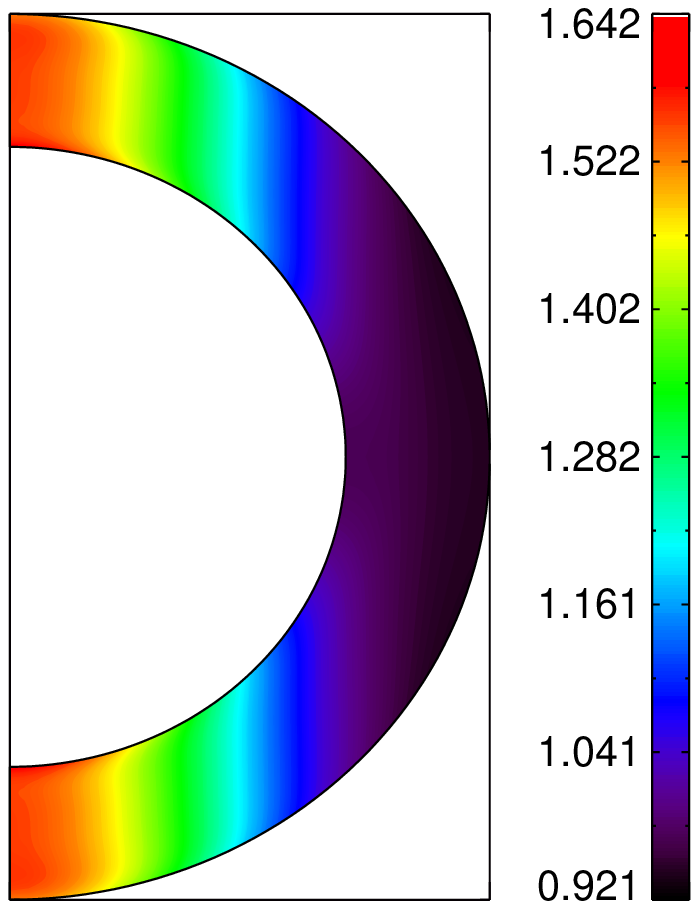}}}
\caption{Top:   Meridional circulation (given as its Reynolds numbers) at the top (blue) and bottom (red) of the convection zone associated with  rotation laws shown in Fig. \ref{fig1}.  Negative  values at the surface of the northern hemisphere indicate a circulation pattern directed towards the poles (counterclockwise flow). Bottom: Similar to Fig. \ref{fig1} but with meridional circulation included. 
Positive $H^{(1)}$ values lead to solar-type acceleration of the equator (left panel).  The (slow) counterclockwise  meridional flow $H^{(1)}= 0$ leads to a weak  antisolar-type equatorial deceleration (middle panel). The right panels of Figs. \ref{fig1}  and \ref{fig2a} formally demonstrate the  possibility of antisolar rotation laws  to produce antisolar rotation laws with negative $H^{(1)}$. }
\label{fig2a}
\end{figure} 

{ The  question remains as to how the meridional circulation   would modifiy these results if it were included.
The above  finding that formally negative $H$ easily produces antisolar rotation profiles  remains true if the meridional circulation  due to radial shear   is  also taken into account.} The vorticity of the circulation depends on the sign of the radial shear $\partial \Om/\partial r$.
At the surface it flows towards the equator for superrotation and towards the poles for subrotation  \citep{K63}.
On the other hand, a circulation which flows towards the equator at the surface of the convection zone (``clockwise flow'') produces  differential rotation with an accelerated equator while it produces a polar vortex if it flows towards the poles (``counterclockwise flow'').
 One takes from Fig. \ref{fig2a} (top) that for all choices of $H^{(1)}$ the meridional circulation flows counterclockwise in the northern hemisphere, reducing the equatorial acceleration (left panel) or amplifying equatorial deceleration (right panel). Indeed, with circulation included, the bottom left  plot of Fig. \ref{fig2a} with $H^{(1)}> 0$ represents a model for convection zones with solar-type rotation laws, while  the right panel of Fig. \ref{fig2a} with  $H^{(1) }<0$ represents an antisolar rotation law. { As the middle plots of  Fig. \ref{fig2a} show, a meridional circulation towards the poles even produces a weak antisolar rotation without any $\Lambda$ effect. Typically, as a result of the Taylor-Proudman theorem the isolines of the angular velocity $\Om$ (with meridional flow included) become cylindrical. This effect appears in all plots of the bottom row of 
  Fig. \ref{fig2a} but it is most prominent for   $H^{(1) }<0$ which even without circulation generates cylinder-like $\Om$-isolines. Simultaneously, for rotation laws with small $\partial \Om/\partial z$ the amplitude of the circulation sinks.}

Several analytical studies of the $\Lambda$ effect  led to positive $H^{(1)}$, that is,   $\cos\theta Q_{\theta\phi}>0$,   
for rigid rotation.
Also,  numerical simulations of rotating convection \citep{HK06} or driven anisotropic turbulence under the influence of solid-body rotation \citep{K19a} provide  transport of angular momentum towards the equator.  Earlier, \cite{C01}  found  transport towards the equator only for fast rotation while for slow rotation occasionally  the opposite result   appeared. Simulations by  \cite{RE05} of  rotating turbulent  convection with much higher resolution    provided  very small  $Q_{\theta\phi}$ for slow rotation and large  positive  $Q_{\theta\phi}$ for fast rotation. There seemed to be no hope, therefore,  of explaining  antisolar rotation laws for  rotating stars with a hydrodynamical theory  of turbulent rotating flows.
In this paper  numerical simulations of rotating convection in boxes   are presented providing transport of  angular momentum towards the poles for slow rotation. This  transport towards the poles, however,     does{ not} result from the $\Lambda$ effect but is   due to the rotation-induced  off-diagonal viscosity   term $\hatnuT$  in (\ref{dr02}) in { connection with a subrotation law  $\partial \Om/\partial r<0$
 which appears for slow rotation \citep{VW18}.
For solar-like convection zones 
with  rotation profiles   quasi-uniform in radius and rotating with the present-day solar rotation rate  the $ \hatnuT$ term does  not play any  role.}


 \section{Rotating convection}\label{Rotating}

 We  perform simulations for convection with a fixed  ordinary Prandtl number $\Prr=\nu/\chi$   with  $\chi$ being the thermal diffusion coefficient.  For stellar material the heat conductivity $\chi$ strongly exceeds  the other  diffusivities.    
 Also the Roberts number $ {\rm q}={\chi}/{\eta}$ with $\eta$ as the microscopic magnetic resistivity is therefore much larger than unity. 
For numerical reasons we must work with the approximate surface   value $\Prr=0.1$.

The  simulations are done with the {\sc Nirvana} code by \cite{Z02}, which uses a conservative finite volume scheme in Cartesian coordinates. The length scale is defined by the depth of the convectively unstable layer.  
Periodic boundary conditions are formulated in the horizontal plane. The upper and lower boundaries are impenetrable and stress-free. 
The initial state is convectively unstable in a layer that occupies half of the box.  Convection sets in if the Rayleigh number
 exceeds its critical value. In the dimensionless units  the size of the simulation box is $2\times6\times6$ in the $x$, $y$, and $z$ directions, respectively. The lower and upper boundaries of the unstably stratified layer are located at $x=0.8$ and $x=1.8$, respectively.
 The numerical resolution is $128\times384\times384$ grid points.
The stratification of density, pressure, and temperature is piecewise polytropic, similar to that used in \cite{RK12}. The density varies by a factor five over the
 depth of the box, hence the density scale height is 1.2.

{The code solves the momentum equation, 
\begin{eqnarray} 
\rho\left( \frac{\partial  \vec{u}}{\partial t} + (\vec{u} \cdot\nabla ) \vec{u} \right) = -\nabla p +
   \nabla \cdot \tau + \rho \vec{g} - 2\rho \vec{\Om} \times \vec{u},
   \label{motion}
\end{eqnarray}
where $\rho$ is the mass density, $\vec{u}$ the gas velocity, $p$ the gas pressure, $\vec{g}$ gravity,  and $\vec{\Om}$ the rotation vector,
in a corotating Cartesian box under  mass conservation, 
\beg
  {\partial \rho}/{\partial t} + \nabla \cdot (\rho \vec{u}) = 0,
  \ende 
   together with  the energy equation, 
\begin{eqnarray}
   \frac{\partial e}{\partial t} + \nabla \cdot \left(\left(e+p\right)\vec{u}  \right)  = 
    \nabla \cdot \left( \vec{u}\cdot \tau  - \vec{F}_{\rm cond} \right),
    \label{energy}
\end{eqnarray}
where $\vec{F}_{\rm cond}=-\kappa \nabla T$  being the conductive heat flux with  the heat conduction coefficient $\kappa$.
 The viscosity   tensor is  
\begin{eqnarray}
  \tau_{ij}=\rho\nu \big(u_{i,j}+u_{j,i} - \frac{2}{3} (\nabla \cdot \vec{u}) \delta_{ij}\big)
,\end{eqnarray}
 and the  total energy 
$
  e=U + \rho \vec{u}^2/2
$
 is the sum of the thermal and  kinetic energy densities. An ideal gas with a constant mean molecular weight $\mu=1$ is considered, hence  
\begin{eqnarray}
  U = \frac{\cal R}{\gamma-1} \rho T 
\end{eqnarray}   
for  the  thermal energy density  with $\cal R$ the gas constant and $\gamma=c_{\rm p}/c_{\rm v}=5/3$.
The gas is  kept at a fixed temperature at the bottom and a fixed heat flux at the top of the simulation box. 
More technical details including  the boundary conditions have been described  in \cite{RK12}.}
 \begin{figure}[hbt]
 \centering
\hbox{
\includegraphics[width=4.4cm]{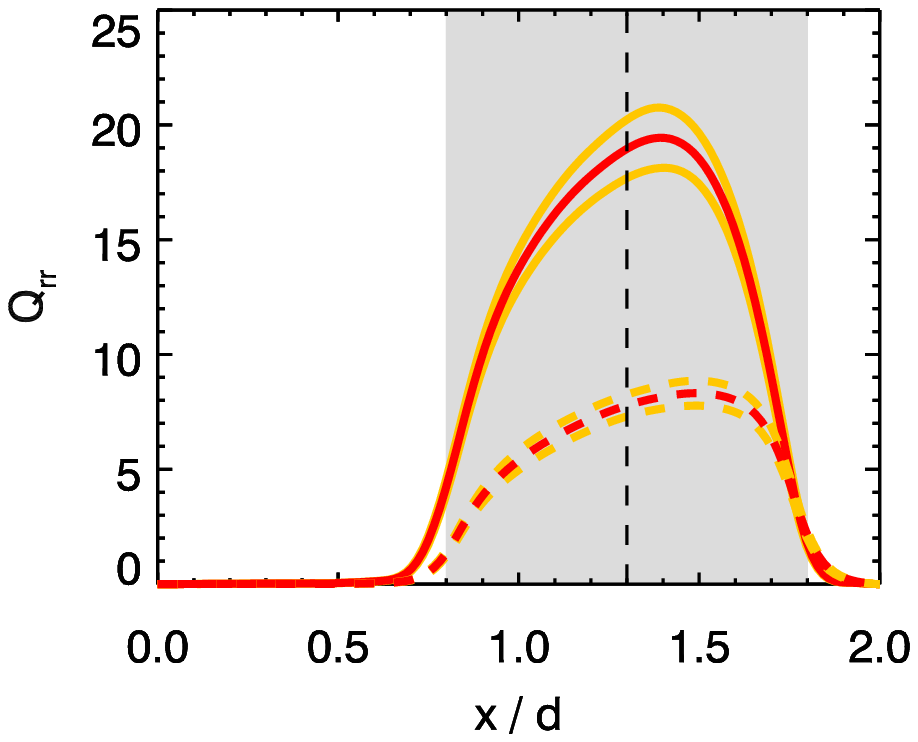}
\includegraphics[width=4.4cm]{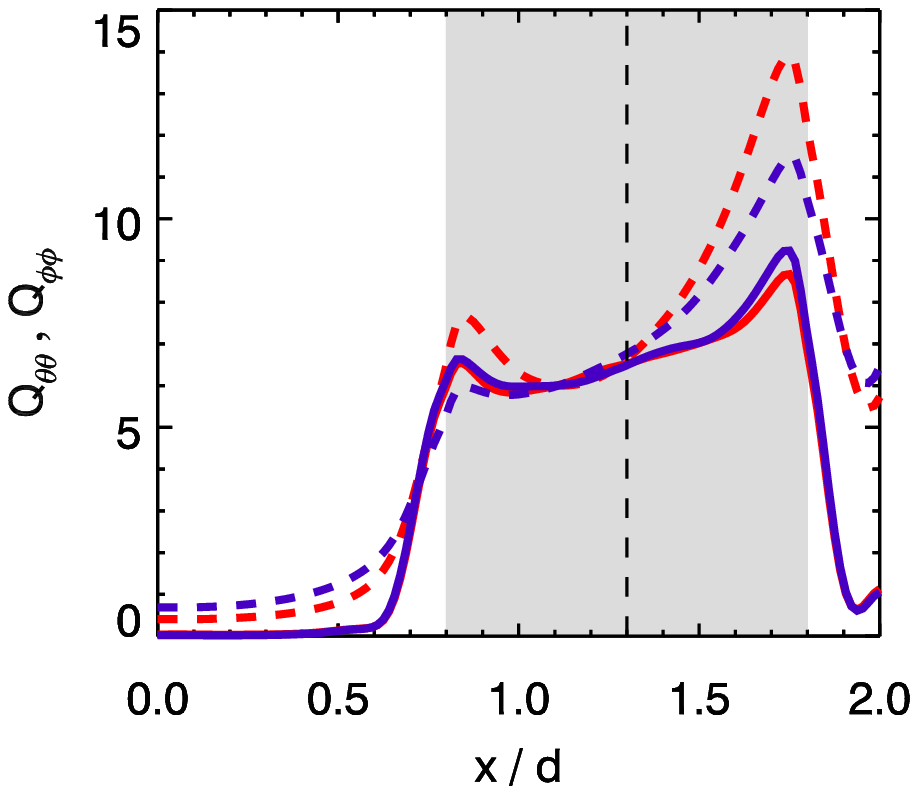}}
\caption{Convection of  very slow rotation   ($\Omst=1$, solid lines) and    very fast rotation   ($\Omst=30$, dashed lines). Left: $Q_{rr}$, right:  $Q_{\theta\theta}$ (red lines)  and  $Q_{\phi\phi}$ (blue lines).
The gray-shaded area indicates the convectively unstable part with the vertical dashed line showing its center and  $d$ is the thickness of the convectively unstable layer. Correlations $ Q_{rr}$, $ Q_{\theta\theta}$,  and $ Q_{\phi\phi}$ and rotation rates $\Om$ are given in code units. The    volume-averaged turbulence intensity is $u^2_{\rm rms}=28$ and the co-latitude $\theta=45^\circ$.}
\label{fig11}
\end{figure} 

Figure \ref{fig11} gives the auto-correlations of the one-point correlation tensor (\ref{Q}) for convection that is subject to slow  and fast rotation. The colatitude is $\theta=45^\circ$.  
As expected, the horizontal turbulent intensities are identical for slow rotation while the vertical intensity $\langle u_r^2\rangle$  has the dominating value. The latter is  strongly suppressed  by faster rotation (the radial turbulence intensity is reduced by more than a factor of two  for $\Omst=30$) while there is almost no  visible  suppression of the horizontal components.

For driven turbulence in a quasilinear approximation we have
 \beg
{Q}_{ij} ={Q}_{ij}^{(0)} - \varepsilon\ (2\Om^2 \delta_{ij}-{\Om_i\Om_j})
\label{Q2}
\ende  
with
\beg
\begin{split}
\varepsilon&=\frac{2}{15}\int^\infty_{0}\int^\infty_{-\infty} \frac{\nu^2 k^4-3\omega^2}{(\omega^2+\nu^2 k^4)^2} E\  {\rm d}k {\rm d}\omega \\\
&= -\frac{2}{15}\int^\infty_{0}\int^\infty_{-\infty} \frac{\omega}{(\omega^2+\nu^2 k^4)^2} \frac{\partial}{\partial \omega}\left((\omega^2+\nu^2 k^4)E\right)\  {\rm d}k {\rm d}\omega
\end{split}
\label{Q3}
\ende  
\citep{R89}.  The sign of the expressions  follows from the form of the positive-definite spectrum $E(k,\omega)$ and is obviously  negative for white-noise spectra but is positive for spectral functions which are sufficiently steep in $\omega$. Also for the maximally steep spectrum such as  $\delta(\omega)$ the $\varepsilon$ is positive,  describing a rotational quenching of the turbulence intensities. 

  The vector of rotation at the colatitude $\theta$ is $\vec{\Om}=\Om(\cos\theta,-\sin\theta,0),$ so that Eq. (\ref{Q2}) gives    
 \beg
\langle u^2_r\rangle =\langle u^{(0)2}_r\rangle -\varepsilon\Om^2(2-\cos^2\theta)
\label{Q4}
.\ende     
The  rotational quenching of the radial turbulence intensity  shown  in Fig. \ref{fig11}   can be described by Eq. (\ref{Q4}) with  $\varepsilon>0$. 
\begin{figure*}[hbt]
 \centering
 \includegraphics[width=4.3cm]{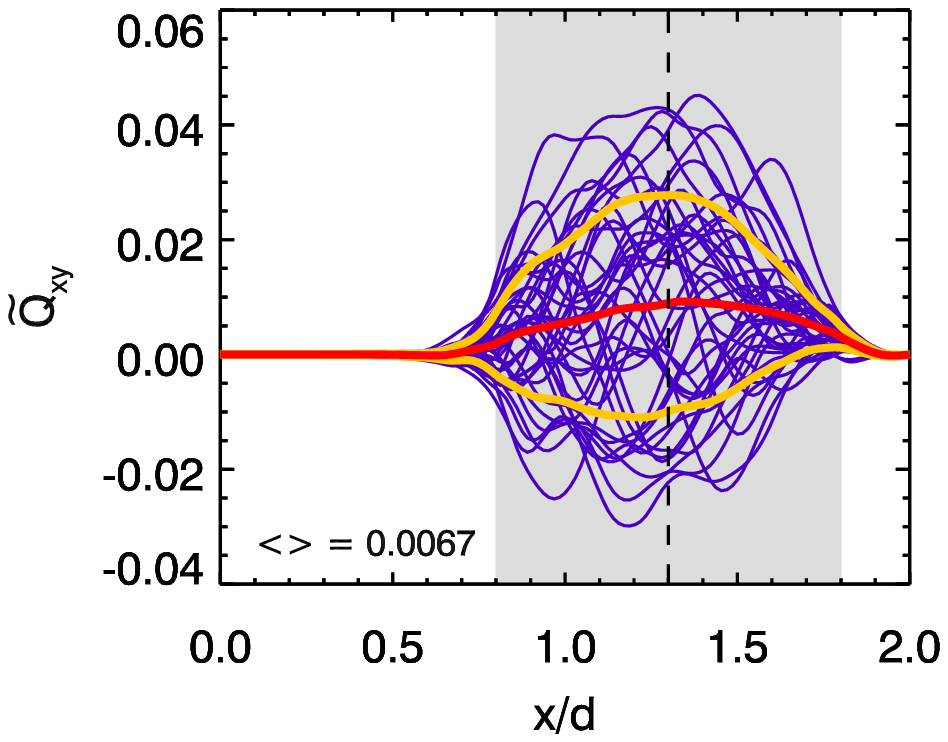}
\includegraphics[width=4.3cm]{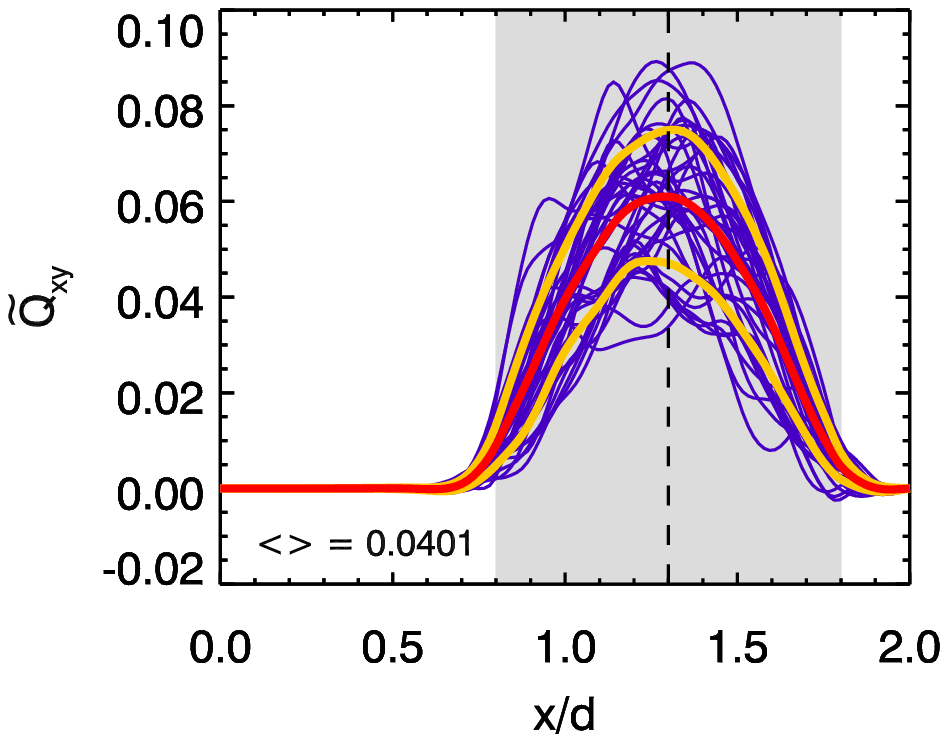}
\includegraphics[width=4.3cm]{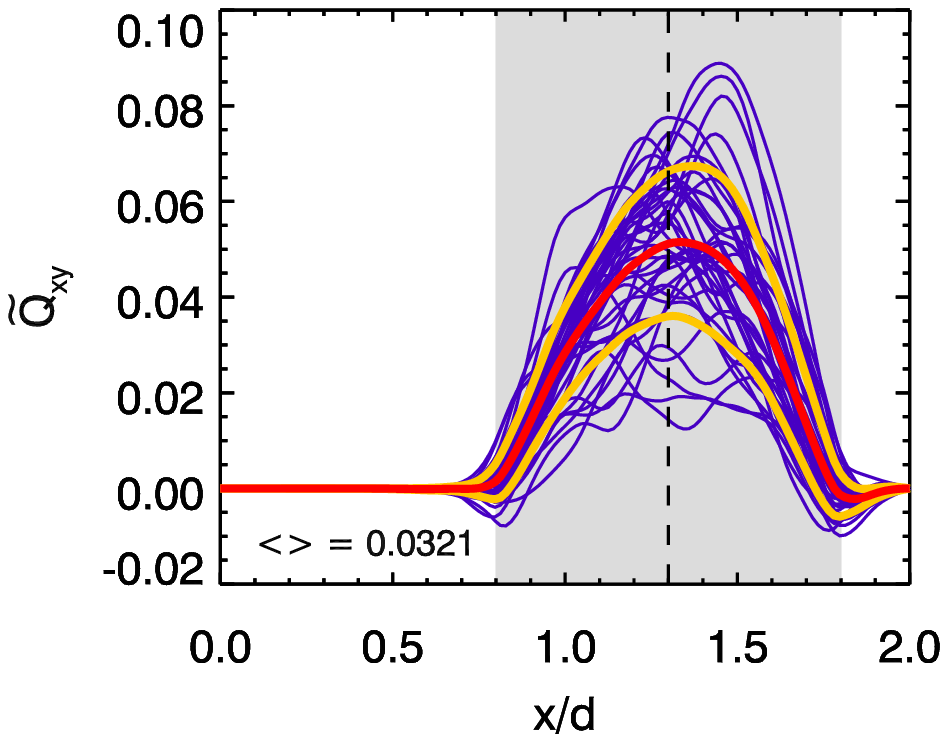}
\includegraphics[width=4.3cm]{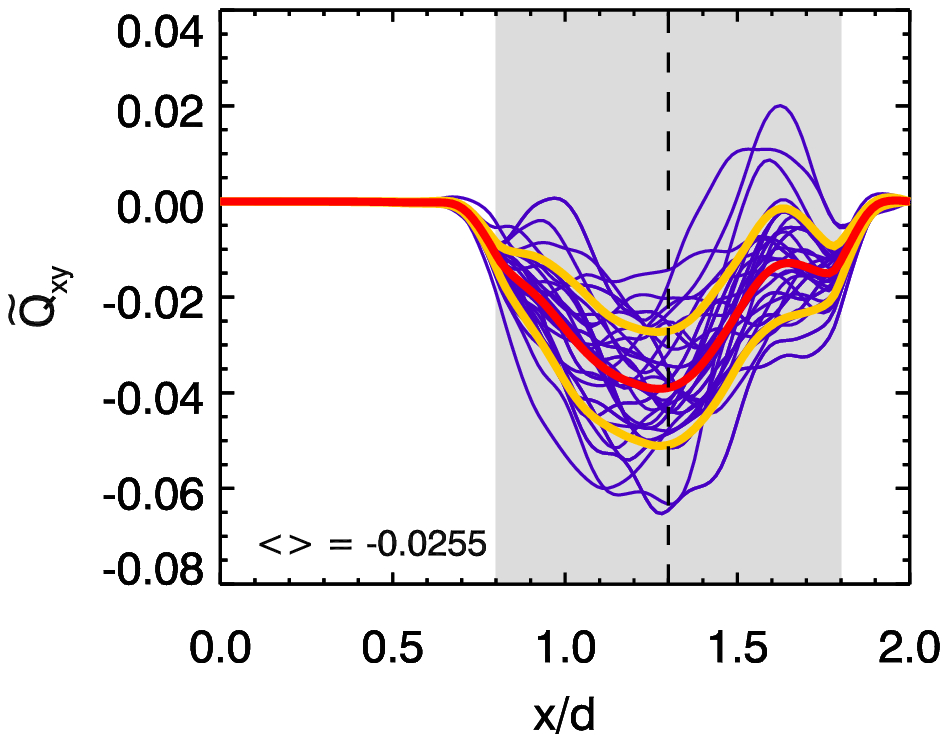}
\caption{Snapshots of the radial profiles of the radial cross-correlation  ${\widetilde Q}_{xy}$ (normalized with the  volume-averaged rms velocity,  $u^2_{\rm rms}$)  for $\Omst=1$, $\Omst=3$, $\Omst=5,$ and $\Omst=10$ (from left to right).
 The  sign changes for $\Omst>5$  from positive  to negative. The convectively unstable part of the box is gray-shaded.
  $\Prr=0.1$, $\theta=45^\circ$.}
\label{figrad}
\end{figure*} 

 Another direct consequence of (\ref{Q2}) is the existence of the cross-correlation of radial and latitudinal fluctuations, that is,
   \beg
\Qrt =-\varepsilon\ \Om^2\ \sin\theta\cos\theta,
\label{Q41}
\ende     
which vanishes at the poles and the equator  by definition. In a sense, the correlation $\Qrt$ mimics the turbulent thermal conductivity tensor. If a radial
temperature gradient exists, a negative cross-correlation $\Qrt$ organizes 
heat transport to the poles resulting in a meridional circulation towards the equator at the surface \citep{REK05}.
\begin{figure*}[htb]
 \centering
 \includegraphics[width=4.3cm]{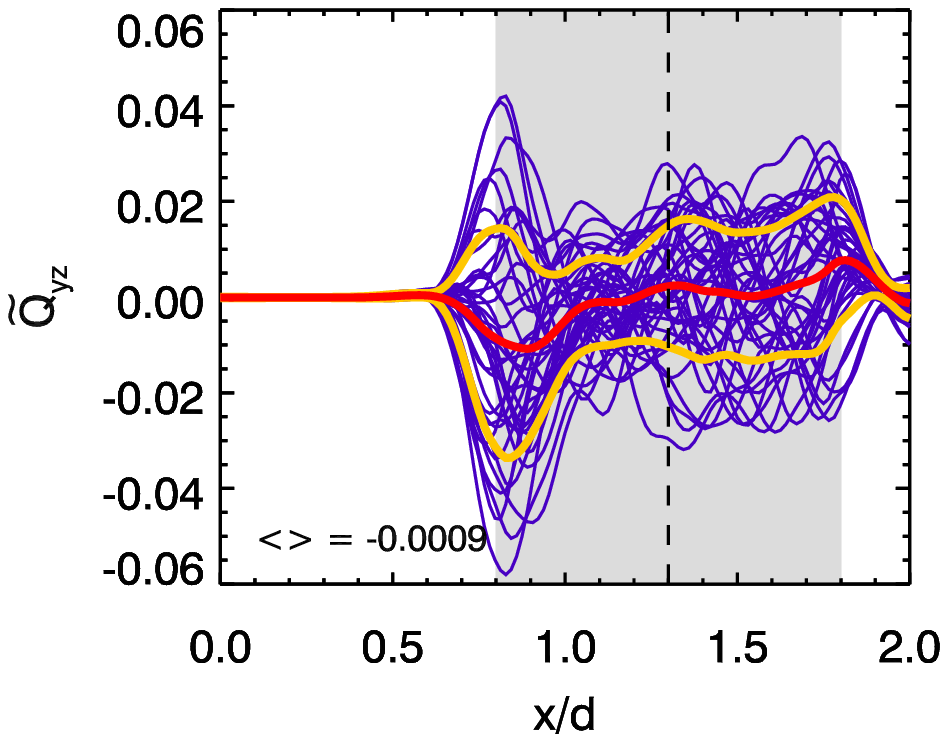}
\includegraphics[width=4.3cm]{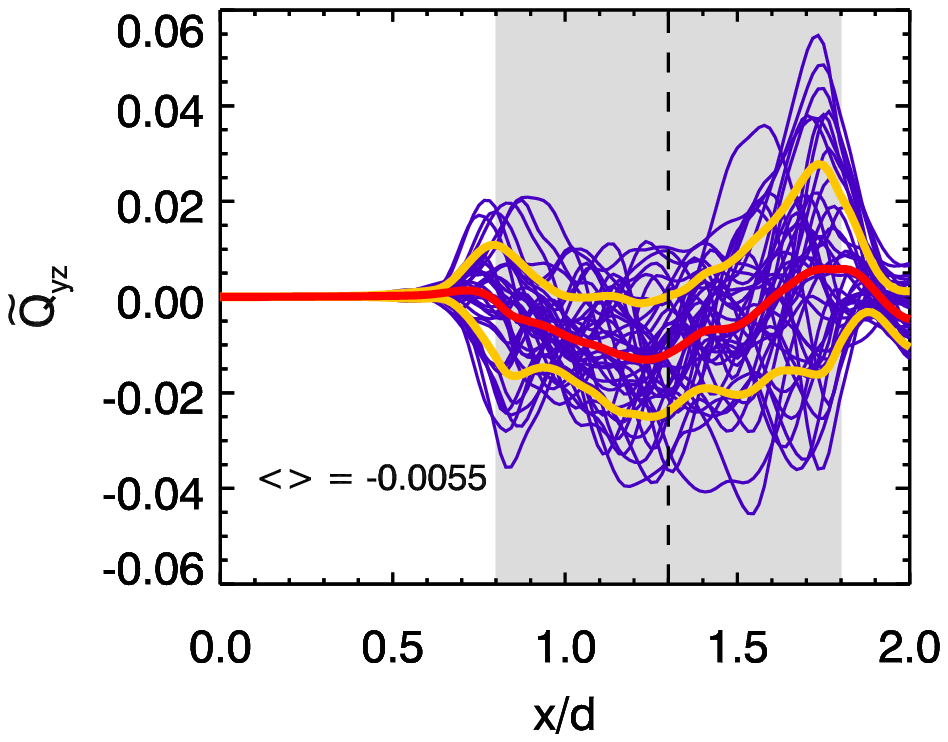}
\includegraphics[width=4.3cm]{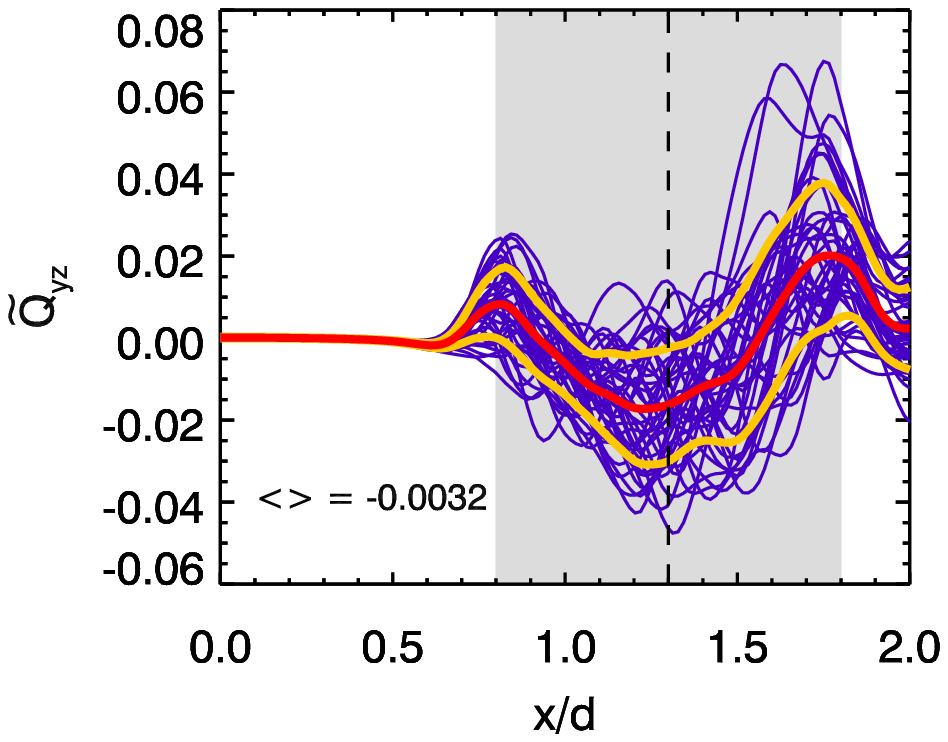}
\includegraphics[width=4.3cm]{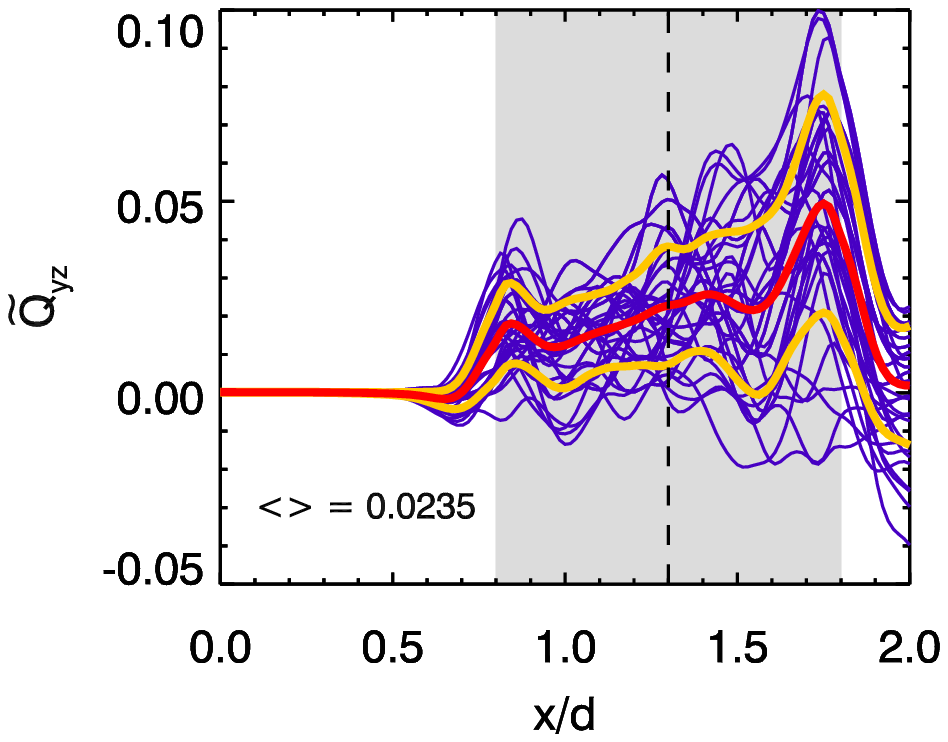}
\caption{Similar to Fig. \ref{figrad} but for the horizontal cross-correlation   ${\widetilde Q}_{yz}$.
The  sign changes for $\Omst>5$  from negative to positive.}
\label{figlat}
\end{figure*}

For positive $\varepsilon,$ the rotation-induced cross-correlation $\Qrt$ after (\ref{Q41}) becomes negative. This theoretical result  complies with  the results of 
numerical simulations  \citep{K19a}.
The expression (\ref{Q2}) only describes
the rotational influence on isotropic turbulence. \cite{RE05} also considered the rotational influence on turbulence fields which are anisotropic in the radial direction 
 with the general result that  $\Qrt$ is always less than zero (northern hemisphere) for steep
spectra $E$ and for all rotation rates.

Hereafter we switch to the Cartesian box coordinates $x$ (representing the radial coordinate  $r$), $y$ (representing the colatitude $\theta$), and $z$ (representing the azimuth $\phi$) hence  $Q_{r\theta}\to Q_{xy}$, $Q_{r\phi}\to Q_{xz}$  and $Q_{\theta\phi}\to Q_{yz}$. The shears $r\partial  \Om/\partial r$ and $\partial        \Om/\partial \theta$  translate into  $\dd U_z/\dd x$  and $\dd U_z/\dd y$, respectively. After averaging over the horizontal  ($yz$) plane  the relations (\ref{dr01})  and (\ref{dr02}) turn into
\beg
Q_{xz}=-\nuT \frac{\dd U_z}{\dd x}+
\nuT {V}\sin\theta \Om
\label{correlation1}
\ende
and
\beg
Q_{yz}= {\hatnuT}\Om^2\cos\theta\sin\theta \frac{\dd U_z}{\dd x}+\nuT{ H} \cos\theta \Om.
\label{correlation2}
\ende 
The cross-correlation (\ref{Q41}) completed by the viscosity term becomes 
 \beg
Q_{xy} =-\varepsilon\Om^2\ \sin\theta\cos\theta- \nuT \frac{\dd  U_y}{\dd x}.
\label{correlation3}
\ende    
The  attention is focused here on the influence of the diffusion terms in Eqs. (\ref{correlation2}) and (\ref{correlation3}) in order to probe the existence of the viscosities $\nuT$ and $\hatnuT$ by simulations. To this end the rotation rates are assumed to be so small that the nondiffusive terms in the relations (\ref{correlation2}) and (\ref{correlation3}) for the cross-correlations are negligible.
The second term in Eq.  (\ref{correlation3}) is positive for outwards decreasing meridional flow $U_y$, for example. Consequently, the cross-correlation $\Qxy$ should be positive  for slow rotation and negative  for rapid rotation,  changing the sign at  a certain value of the parameter $\Omst$ (which denotes the angular velocity $\Om$ of the rotation  in code units). One  finds such a transition from positive to negative values  for $5<\Omst<10$ in the simulations given in  Fig. \ref{figrad}. The coincidence suggests that indeed the influence of viscosity terms in the expressions of cross-correlations may lead to
direction reversals of transport terms as a function of rotation.

Without rotation, all  cross-correlations   vanish. For the slow-rotation models with $\Omst=1,$ already finite values appear  (left  plots in Figs. \ref{figrad} and \ref{figlat}). At the radial boundaries the correlations $Q_{xy}$  and $Q_{xz}$ vanish by the boundary condition  ($u_x=0$) but the   horizontal cross-correlation  $Q_{yz}$ remains finite; it is always positive at the top and bottom of the unstable box  which  indicates $H>0$ if a possible mean circulation $U_z$ were maximal or minimal  at the top or bottom of the convection box (as it is, see Fig. \ref{fig51}).

The correlation $\Qxy$  for slow rotation is  positive so that heat is transported towards the  equator.   At the same time the horizontal  correlation $Q_{yz}$  assumes  negative values.
Independent
of the rotation rates and for both hemispheres we find the general
result that always $\Qxy\Qyz<0$. The simulations therefore show that 
angular momentum flux to the equator (poles) is always accompanied
by heat transport to the poles (equator). Somewhere
between $\Omst=5$  and $\Omst=10$ the cross-correlations $\Qxy$ and
$\Qyz$ change their signs becoming negative ($\Qxy$) and positive ($\Qyz$) for fast rotation.
These signs  are well-known from the
analytical expressions derived for driven turbulence for fast rotation.
In this case, the angular momentum  is transported inward as well as toward
the equator by the convection; in other words, it flows  along cylindric surfaces. 


For increasingly fast  rotation
the amplitudes  of the negative $\Qxy$ are increasing, contrary to
  $\Qyz$ which decrease. This is a basic difference
for the two   cross-correlations. We note that for the transformation
$\Om\to -\Om$ the correlations $\Qyz$  change their sign which is not the
case for $\Qxy$. 
The reason is that $\Qxy$ is even in the rotational
rate $\Om$ while the horizontal  cross-correlation  $\Qyz$  is odd.

\section{The eddy viscosities}\label{Eddy}
\begin{figure*}[htb]
 \centering
\hbox {\includegraphics[width=4.3cm]{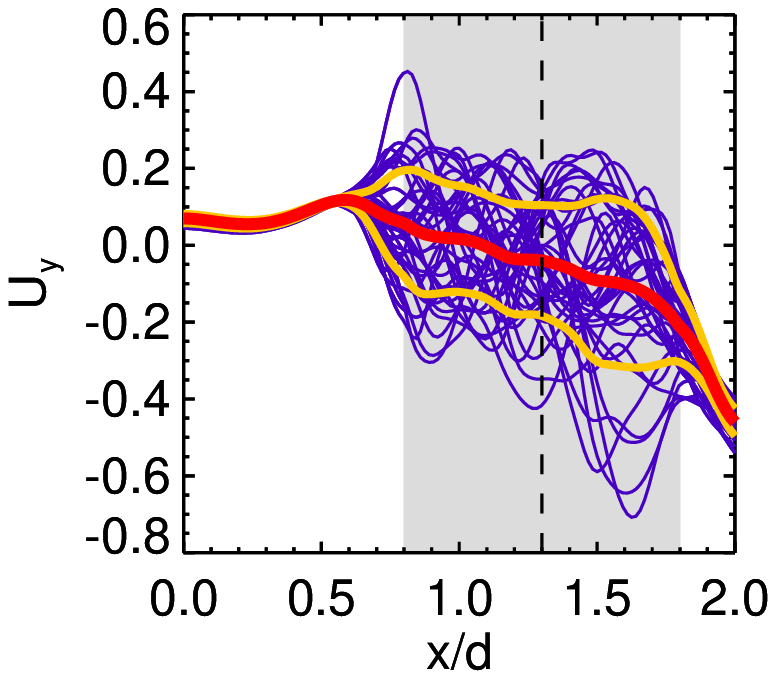}
\includegraphics[width=4.3cm]{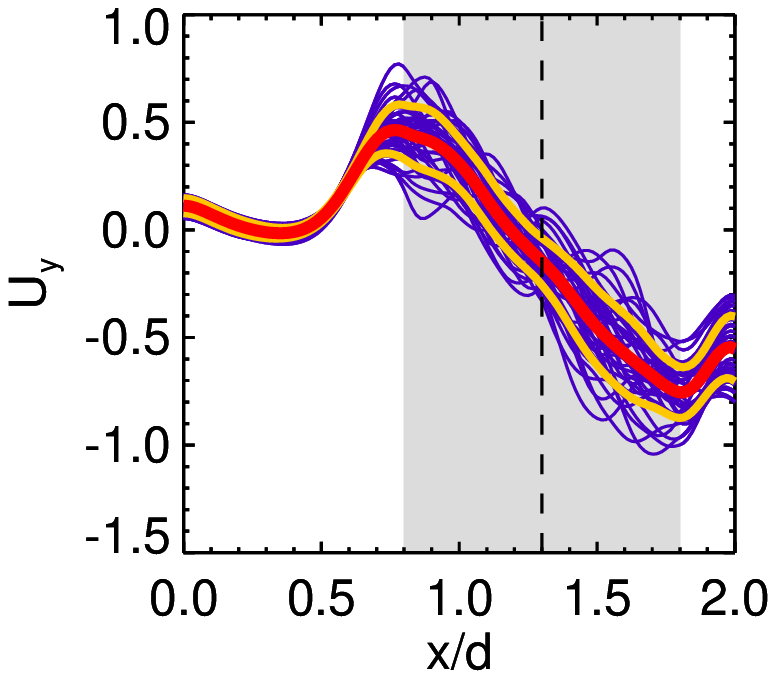}
\includegraphics[width=4.3cm]{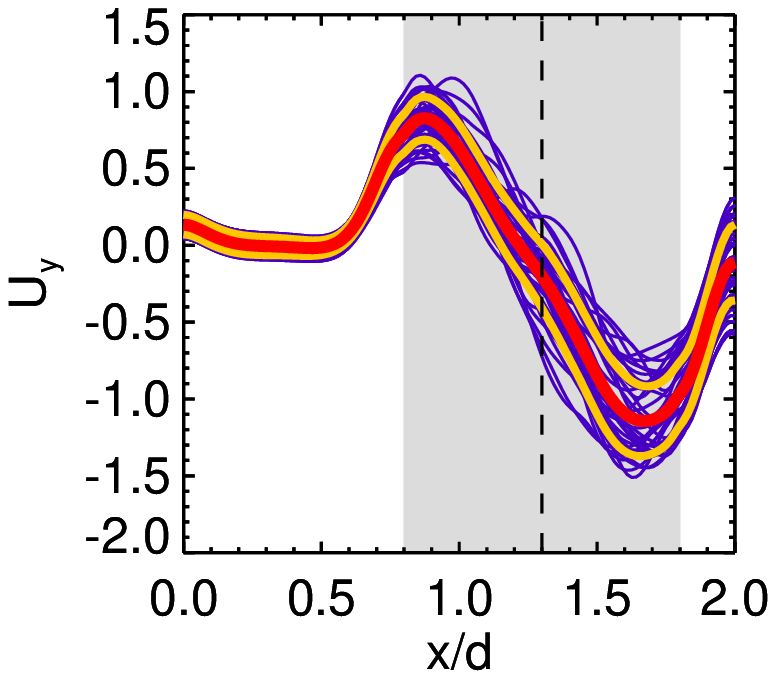}
\includegraphics[width=4.3cm]{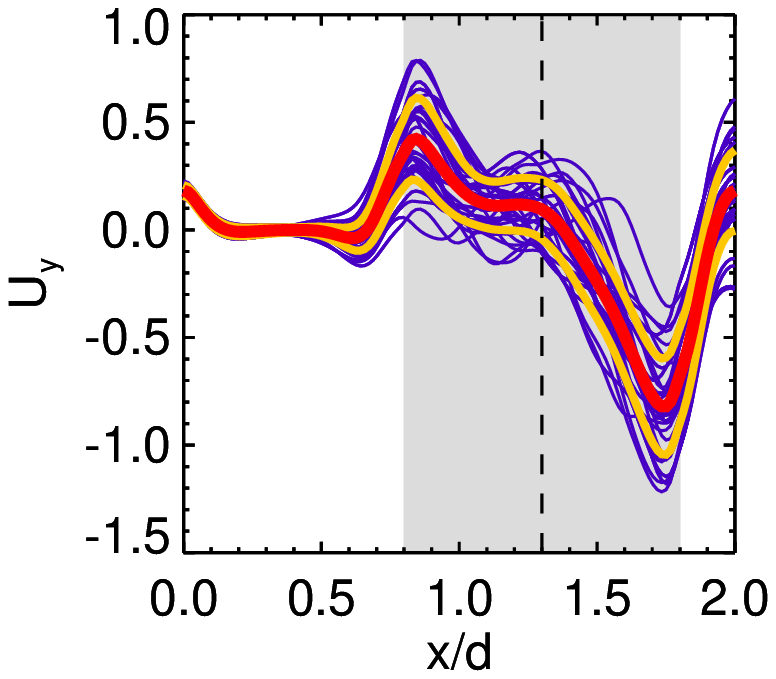}}
\hbox {\includegraphics[width=4.3cm]{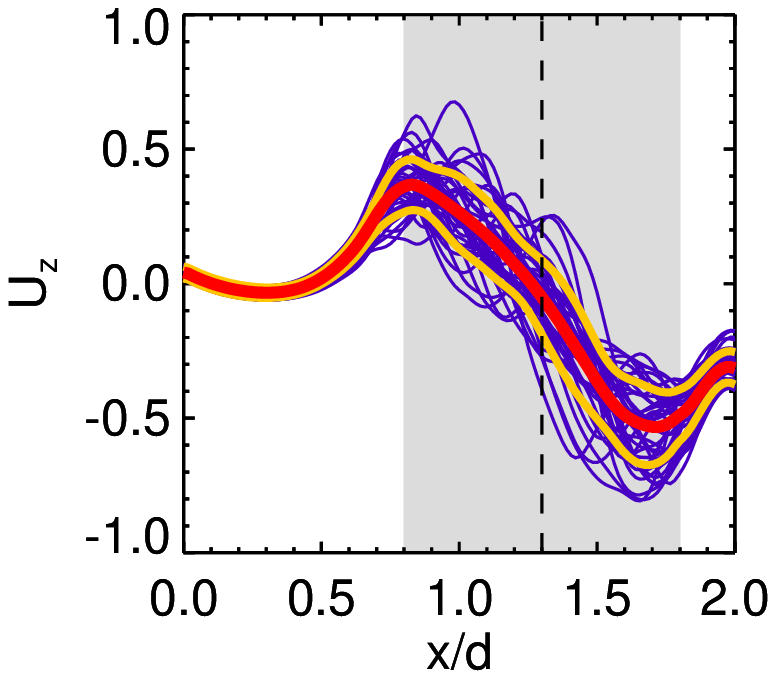}
\includegraphics[width=4.3cm]{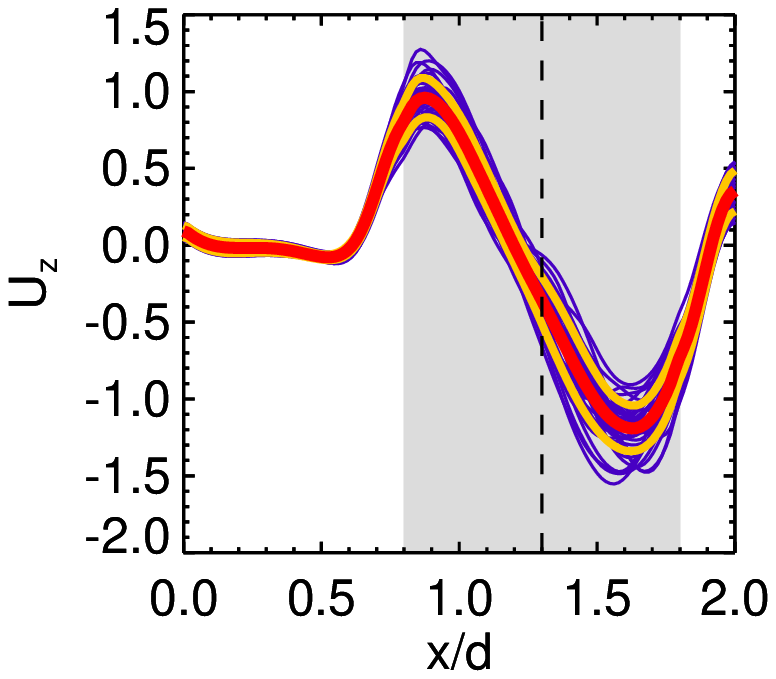}
\includegraphics[width=4.3cm]{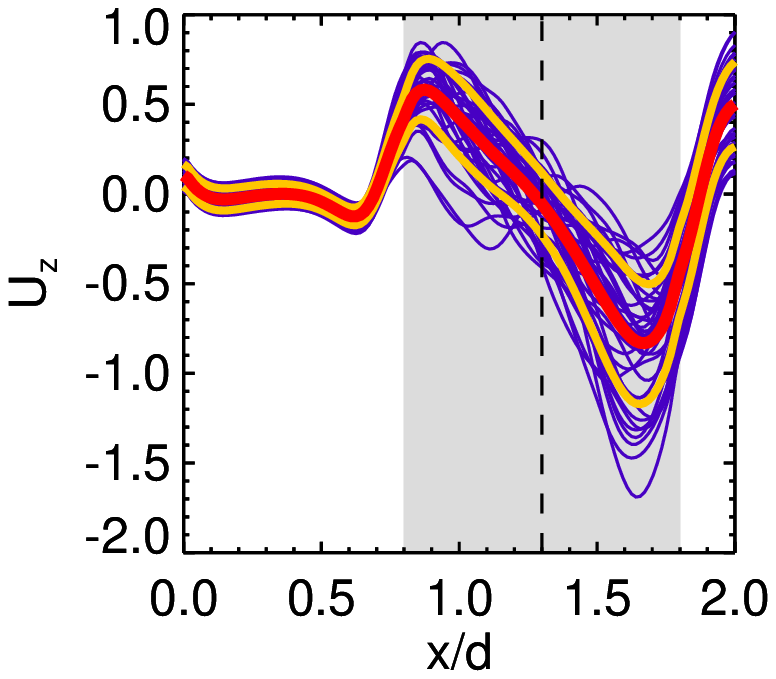}
\includegraphics[width=4.3cm]{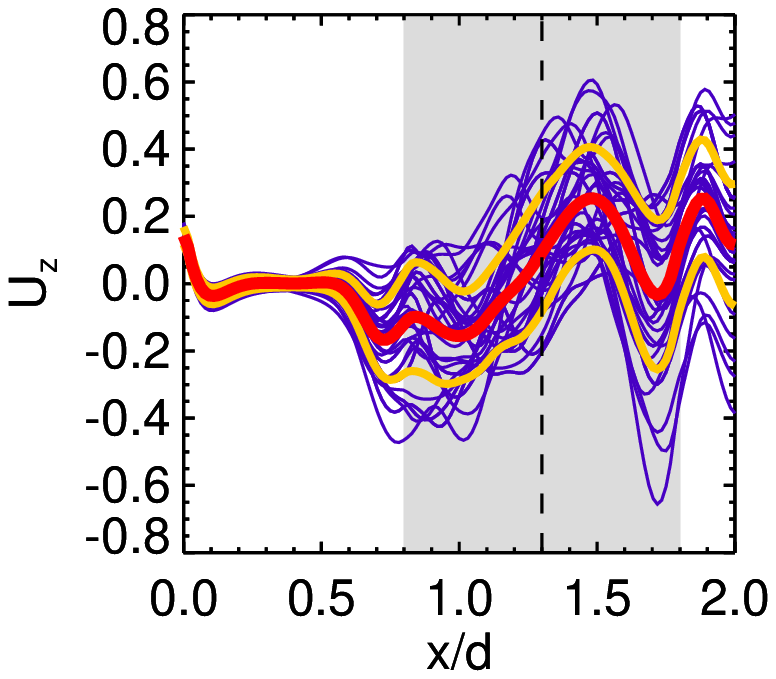}}
\caption{Radial profiles of the meridional flow $U_y$ (top) and zonal flow $U_z$ (bottom) in the convective box (shaded).  The parameters are  $\Omst=1$,  $\Omst=3$,   $\Omst=5,$  and $\Omst=10$  (from left to right).  $\Prr=0.1$. The circulation is always counterclockwise and the  rotation is  decelerated at the surface (subrotation).  $\theta=45^\circ$.}
\label{fig51}
\end{figure*}

Equation (\ref{Q41}) neglects the influence of a  possible radial shear $\dd U_y/\dd x$ of a meridional  flow. 
  The question is whether large-scale mean    flow characterizes the simulation box as in the simulations of \cite{C01} and \cite{KK04},  which  could   be used to calculate the eddy viscosities  after relations (\ref{correlation2}) and (\ref{correlation3}).
The mean flows in the box  have been calculated  for slow rotation. The top panel  of Fig. \ref{fig51} gives   flows in the meridional direction and the bottom  panel gives  zonal flows in the azimuthal direction. The basic rotation there  has been varied from $\Omst=1$ to $\Omst=10$ and the Prandtl number is fixed at  $\Prr=0.1$. The following estimate concerns the first example with $\Omst=1$ with the  cross-correlation  $Q_{xy}\simeq 0.01 u^2_{\rm rms}$ (from Fig. \ref{figlat})  and   the shear $\delta U_y/\delta x\simeq -0.2$ (from Fig. \ref{fig51}). 
The standard eddy viscosity $\nuT$ in code units is 1.5  for slow rotation. The microscopic viscosity of the model in the same units is $\nu=6\cdot 10^{-3}$, meaning that for the Reynolds number $\nuT/\nu\simeq 240$. One  finds   very similar   values for $\Omst=3$. 

The dimensionless eddy viscosity $\alpha_{\rm vis}$, following
 \beg
\nuT=\alpha_{\rm vis} \tau_{\rm corr} u^2_{\rm rms},
 \label{nuT} 
 \ende
may also be introduced which  is often  assumed in turbulence research to be $\alpha_{\rm vis}\simeq 0.3$.
To find the  correlation time $\tau_{\rm corr}$ an auto-correlation analysis as done by   \cite{KR18} is necessary. The result is  $\tau_{\rm corr}\simeq 0.1$ in code units,  hence $\alpha_{\rm vis}\lsim 0.5$ which indeed is of the expected order of magnitude.

Figure  \ref{fig51} also demonstrates   that the negative shear $\dd U_y/\dd x$ is always accompanied by a negative shear $\dd U_z/\dd x$ of the zonal flow, hence $\dd U_y/\dd x\cdot  \dd U_z/\dd x>0$. Transformed to the  global system, this means that a subrotating shell generates  a counterclockwise circulation where the fluid drifts towards the poles at the surface. This type of flow pattern has already been described  in the text below Eq. (\ref{dr12}). On the other hand, the existence of $U_z$ allows us to estimate the off-diagonal viscosity  after Eq. (\ref{correlation2}) with $H\approx 0$ for $\Omst=1$ or $\Omst=3$ to $\hatnuT\simeq 0.05$,   hence $\hatnuT/\nuT\simeq 0.03$ in code units or  $\hatnuT/\nuT\simeq 3 \tau^2_{\rm corr}$ in physical units. 

Following Eqs. (\ref{correlation1}) and (\ref{correlation2}),  the shear ${\dd U_z}/{\dd x}$ contributes to the  cross-correlations $Q_{xz}$ and $Q_{yz}$.  Because of its negativity, the viscosity term in (\ref{correlation1}) is positive, meaning that the actual
value of $|V|$ is even larger than indicated by $Q_{xz}$ from the
simulations.
\subsection{Rotation-induced off-diagonal eddy viscosity}
The same radial velocity gradient  appears in the expression for the horizontal cross-correlation
 in a higher order of the rotation rate. 
 The two viscosities in  (\ref{correlation2}) can be expressed by the integrals
 \beg
 \nuT=\frac{2}{15}\int^\infty_{0}\int^\infty_{-\infty} \frac{\nu^3 k^6}{(\omega^2+\nu^2 k^4)^2} E\  {\rm d}k {\rm d}\omega
\label{Q30}
\ende  
and
\beg
\hatnuT=\frac{48}{105}\int^\infty_{0}\int^\infty_{-\infty} \frac{\nu k^2\omega^2(5\nu^2 k^4-3\omega^2)}{(\omega^2+\nu^2 k^4)^4} E\  {\rm d}k {\rm d}\omega
\label{Q29}
\ende  
\citep{R89}. Not surprisingly, the first integral is positive definite. 
On the other hand, the second integral  is positive for all other monotonously decreasing spectra in line with the following expression:  
\beg
\begin{split}
\hatnuT=\frac{48}{105}\int^\infty_{0}\int^\infty_{-\infty} &\frac{\nu k^2\omega^2}{(\omega^2+\nu^2 k^4)^2}  \\\ &
\left(    \frac{E}{\omega^2+\nu^2 k^4} -2 \frac{\partial}{\partial\omega} \frac{\omega E}{\omega^2+\nu^2 k^4}  
  \right)
   {\rm d}k {\rm d}\omega.
\end{split} 
\ende
The  integral in (\ref{Q29}) is even  positive for spectra of the  white-noise-type. To  demonstrate this point, we evaluate 
the frequency integral   with  uniform $E$. As
\beg
\int^\infty_{-\infty} \frac{\nu^3 k^6\omega^2(5\nu^2 k^4-3\omega^2)}{(\omega^2+\nu^2 k^4)^4}  {\rm d}\omega=   \frac{\pi}{8}
\label{Q31}
,\ende  
here the $\hatnuT$ is also  positive.
We note however that very steep spectra such as $\delta(\omega)$  lead  to   vanishing $\hatnuT$ which explains the absence of antisolar rotation laws in the calculations based on that turbulence model   \citep{KP94}. It therefore seems likely that  observations  of  slowly rotating stars  with a decelerated equator question the application  of turbulences with $\delta$-like frequency  spectra in  stellar convection models. The steepest frequency  spectra describe fluids in the  high-viscosity limit while  spectra with a quasi-white-noise behavior belong to the inviscid approximation.  

As the corresponding integral for the standard eddy viscosity  (\ref{Q30}) is $\pi/2$ one obtains, for rather flat  spectra,  $\hatnuT/\nuT\ \simeq  ( l^2_{\rm corr}/2 \nu)^2\simeq  \tau_{\rm corr}^2$.
 The latter relation requires that    for  the background viscosity $\nu$ the standard expression $\nu\simeq 0.5  l^2_{\rm corr}/\tau_{\rm corr}$ be used. 
Because of $\hatnuT>0$ and  ${\dd U_z}/{\dd x}<0$ (see   Fig. \ref{fig51})   { negative } contributions to the horizontal cross-correlation are produced. The existence of the off-diagonal viscosity $\hatnuT$ therefore  explains the resulting negativity of the cross-correlation $Q_{yz}$ for slow rotation (see Fig. \ref{figlat}).  The  result confirms  the above finding that all values of  $Q_{yz}$ are positive at the top and bottom  boundaries where the mean shear vanishes. For faster rotation, the increasing positive values of $H$  overcompensate the negative contribution from the radial shear  of $U_z$ which becomes  increasingly unimportant  (see Fig. \ref{fig51}). 
The transition from negative to positive $Q_{yz}$ happens in Fig. \ref{figlat}  for $\Omst\simeq 5$. 
Figure~\ref{fig51} also demonstrates that the shear $\dd U_z/\dd x$ grows for  $\Omst<5$  and overcompensates the positive $H$ term producing the obtained  negative cross-correlations.

As the mechanisms of the null crossings of $\Qxy$ and $\Qyz$ are different, the values for the critical $\Omst$ should not be identical for both cases.
\subsection{Antisolar rotation}\label{antisolar}
\begin{figure}[htb]
 \begin{center}
\hbox{
\includegraphics[width=2.75cm]{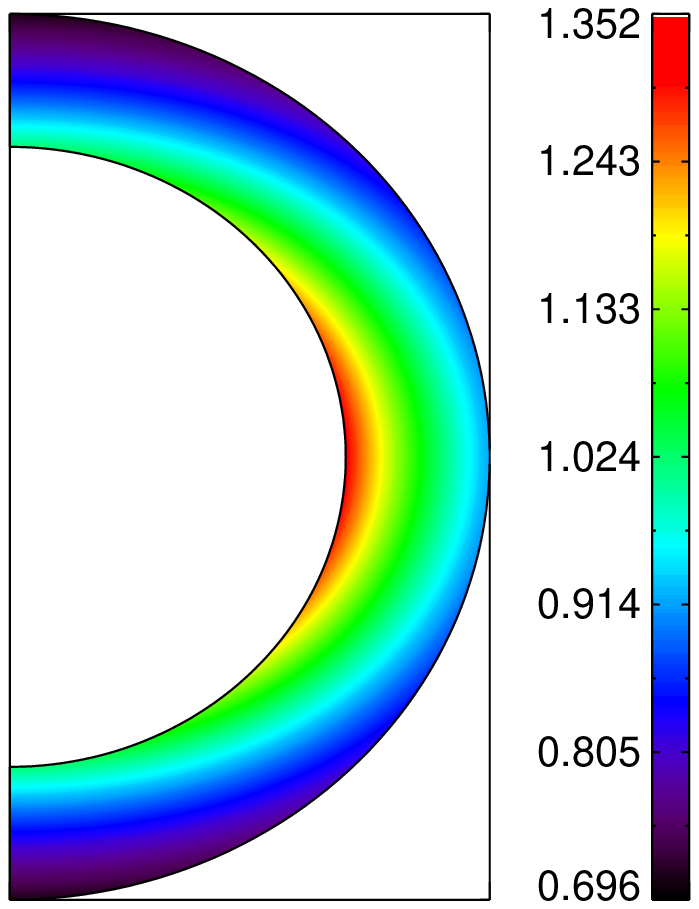}
\includegraphics[width=2.75cm]{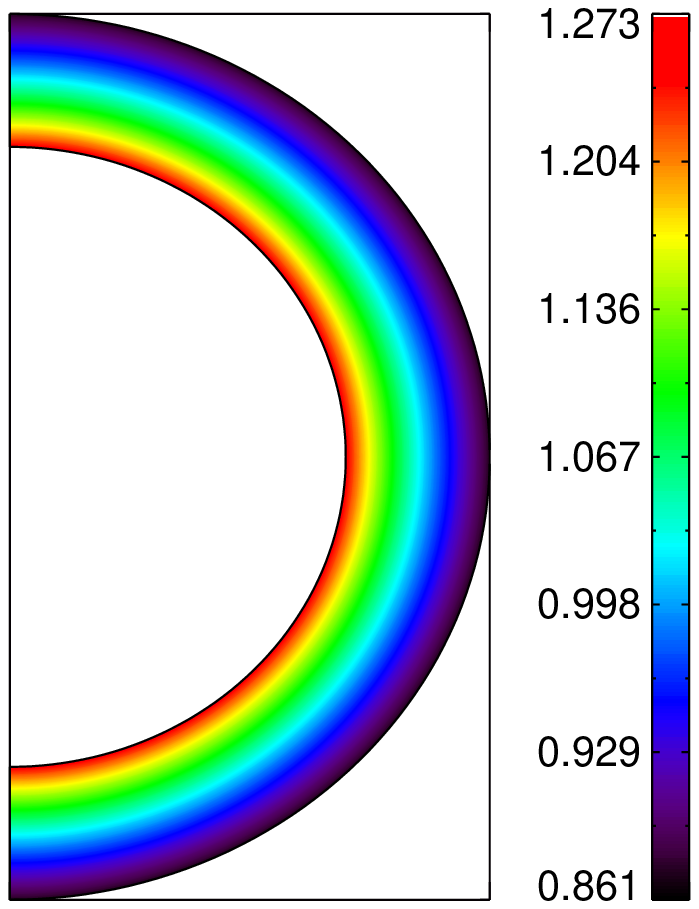}
\includegraphics[width=2.75cm]{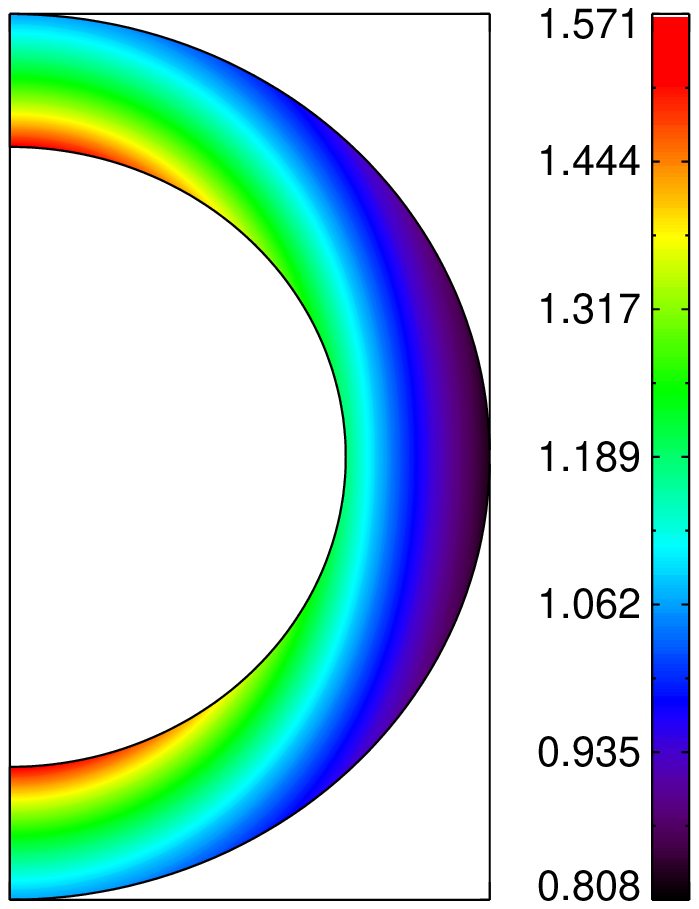}
}
\hbox{
\includegraphics[width=2.75cm]{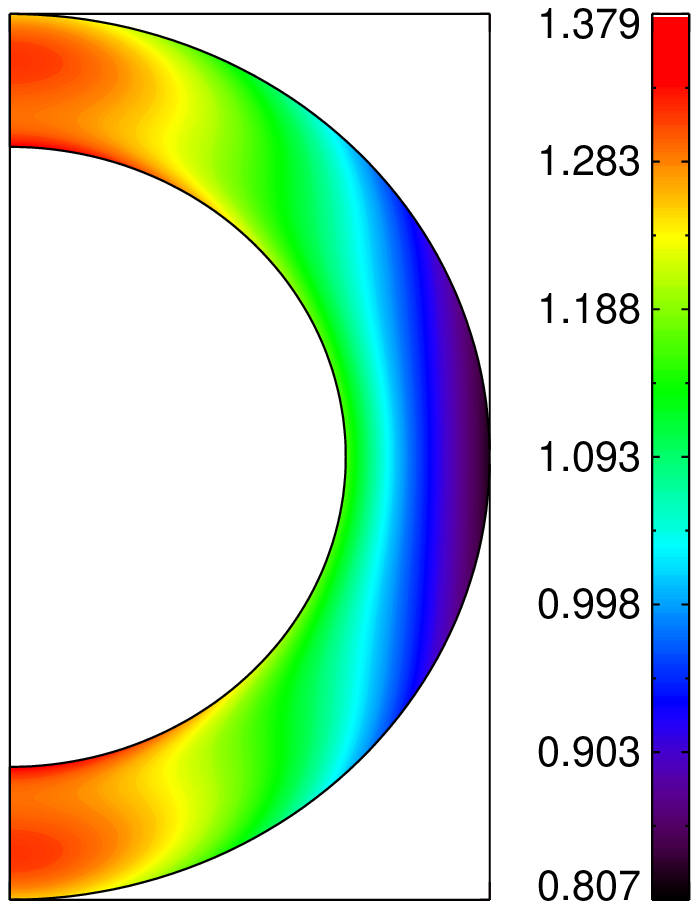}
\includegraphics[width=2.75cm]{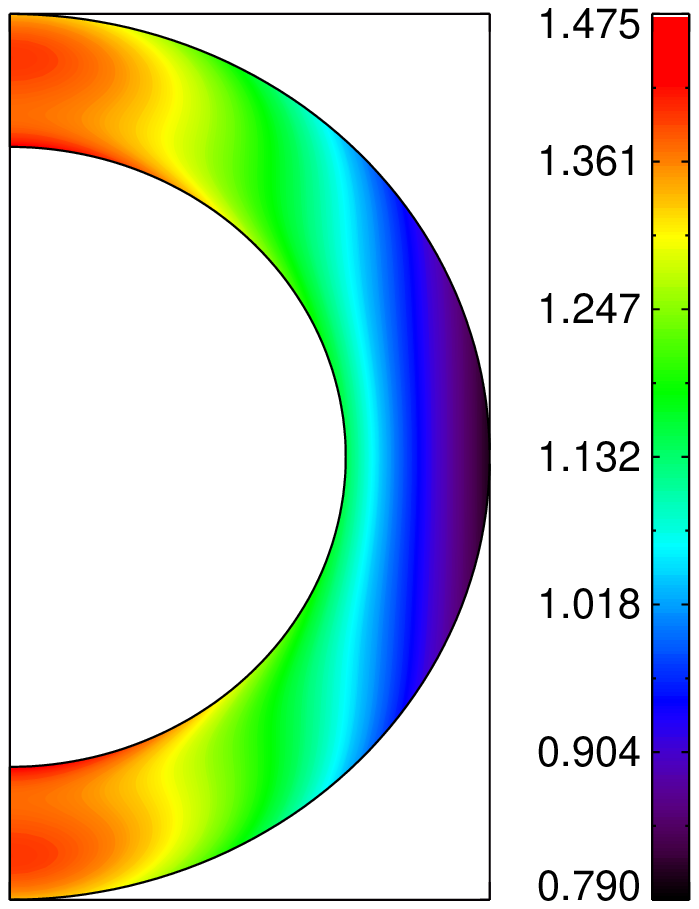}
\includegraphics[width=2.75cm]{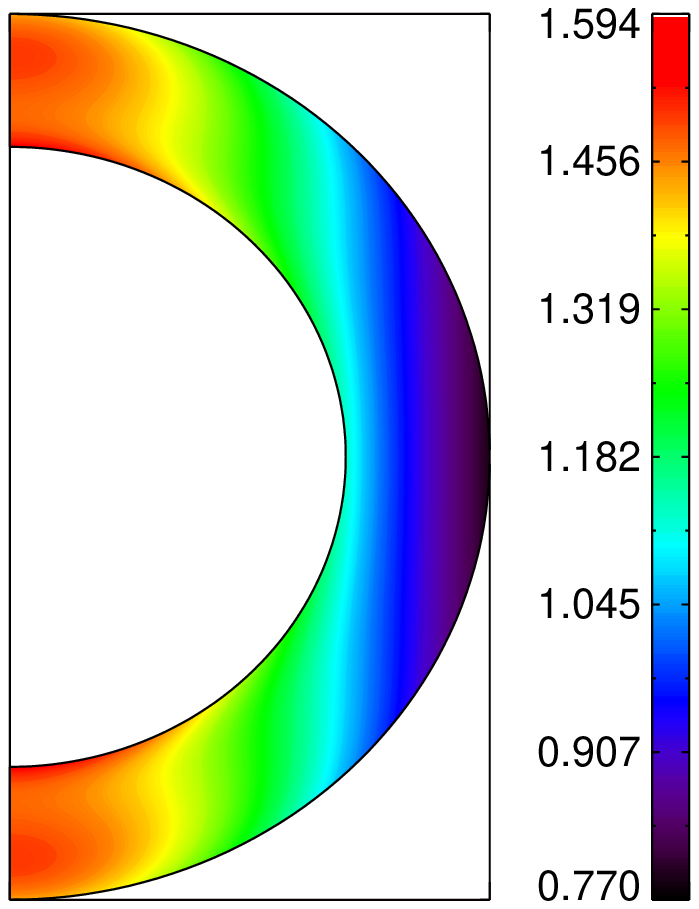}
}
\end{center}
 \caption{Similar to Fig. \ref{fig1}  but for slow rotation  ($V^{(0)}=-1$ and $V^{(1)}=H^{(1)}=0$). The off-diagonal viscosity term varies from  $\hatnuT=-0.5$ (left), through  $\hatnuT=0$ (middle), to  $\hatnuT=0.5$ (right). Meridional circulation is  suppressed (top) or it   is included as in Fig. \ref{fig2a} (bottom).  For positive $\hatnuT$ the rotation is always antisolar without and with meridional circulation.
The circulation is always counterclockwise.
We note the negative radial shear below the equator in all cases.}
\label{fig20}
\end{figure}

 Models are considered of  such slow rotation that $H\simeq 0$ and a (negative) $V^{(0)}$ exists in addition to  the rotation-induced off-diagonal viscosity ${\hatnuT}$. The existence of the latter has been indicated  by the simulations for the horizontal cross-correlation $\Qyz$ shown in Fig. \ref{figlat}. This leads to functions of $H$  that are formally negative (for slow rotation) for which Fig. \ref{fig1} demonstrates the appearance of antisolar rotation laws. The off-diagonal element $\hatnuT $ is varied  in Fig. \ref{fig20} from ${\hatnuT}=-0.5$ (left  panel) through zero (middle panel)  to  ${\hatnuT}=0.5$ (right panel). For the models shown in the top row of the plots the resulting meridional circulation is artificially  suppressed.  One finds that 
 negative $V^{(0)}$  always produces rotation laws with negative radial shear, which
 in combination with  
 positive ${\hatnuT}$ leads to antisolar rotation and vice versa. Hence,   for $V^{(0)}{\hatnuT}<0$ the equator  rotates slower than the polar regions and just this condition is the result of  the given numerical simulations.  One could also
 demonstrate that the equator is accelerated for  $V^{(0)}{\hatnuT}>0$ (not shown).

If the associated meridional flow is also allowed to transport angular momentum, as done in the models in the second row of Fig. \ref{fig20}, the $\Om$ isolines become cylindrical in accordance with the Taylor-Proudman theorem and  the antisolar rotation law is only modified but not destroyed. We note how in the middle panels (only)  the meridional circulation changes the type of the rotation law from uniform on spherical shells to cylindrical with respect to the rotation axis. The circulation cells always flow counterclockwise, that is, towards the poles at the surface.

\begin{figure}[htb]
\begin{center}
\includegraphics[width=\columnwidth]{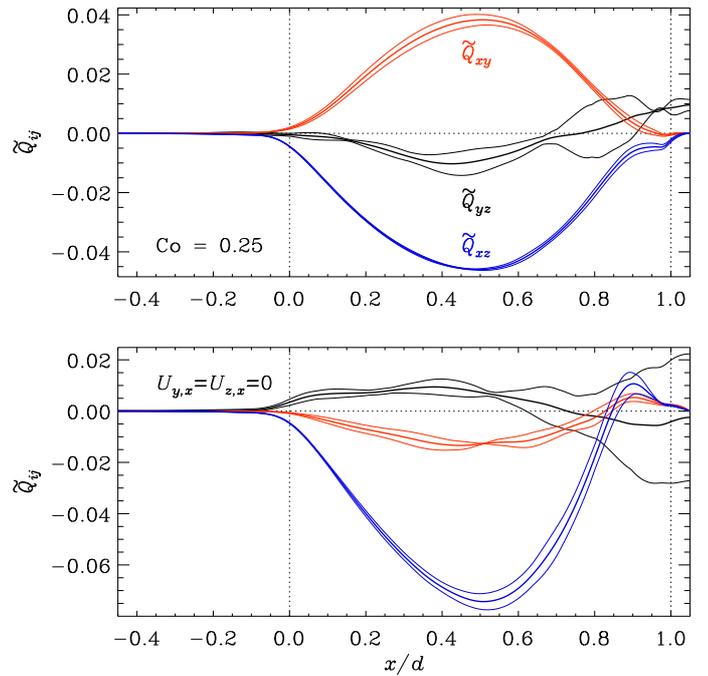}
\end{center}
\caption{Normalized off-diagonal Reynolds stresses as indicated by the
  legends from runs with (upper panel) and without (below) horizontal
  mean flows.  The vertical dotted lines
  indicate the  top and bottom of the convection zone.}
\label{fig21}
\end{figure}

From the comparison of Figs.~\ref{fig1} and \ref{fig20} one finds that
antisolar rotation profiles result both for positive $\hatnuT$ in
common with subrotation and/or for negative $H^{(1)}$. The latter can
be excluded with numerical experiments where $U_y$ and $U_z$ are
artificially suppressed mimicking the existence of strict solid-body
rotation. This has been confirmed by {\sc Pencil
  Code}\footnote{\url{http://github.com/pencil-code}} simulations
which use the same setup as in \cite{K19}; see Fig.~\ref{fig21}.

Here we present results from a slowly rotating run 
with and without mean flows.  The used Coriolis
number ${\rm Co}=\Om d/\pi u_{\rm rms}\approx0.25$  corresponds to $\Omst\simeq 4$ in  {\sc Nirvana} code units. We find that the signs of $Q_{xy}$ and
$Q_{yz}$ change when the mean flows are suppressed. This is consistent
with a dominating contribution from turbulent viscosity in $Q_{xy}$
and the rotation-induced off-diagonal viscosity term in $Q_{yz}$ in
accordance with theory. Finally, we note that the magnitude of
$Q_{xz}$ also increases clearly in the case where the shear is
suppressed, indicating a strong contribution from the term involving
$\nuT$.

\section{Discussion}\label{Discussion}
 The cross-correlations $\Qrt$, $\Qrf$, and $\Qtf$ of the  fluctuating velocities in a rotating turbulence  are playing basic roles in   understanding the rotation laws of stars with outer convection zones. Briefly, the tensor component  $\Qrt$ transports thermal energy in the latitudinal direction ("warm poles") while   $\Qrf$ and $\Qtf$ transport angular momentum in the radial  and  meridional directions. For driven turbulence of a uniformly rotating density-stratified medium (stratified in the radial direction) the correlations fulfill simple sign rules independent of the rotation rate: it is  $\Qrt<0$, $\Qrf<0$ and  $\Qtf>0$, taken always in the northern hemisphere. In physical quantities this means that thermal energy is transported to the poles while the  angular momentum is transported inward and towards the equator. The resulting warm poles drive a clockwise meridional circulation (northern hemisphere) which together with the positive $\Qtf$ transports angular momentum towards the equator.  
 Hence, if the convection zone can be modelled by  driven turbulence under  fast global rotation then the resulting surface rotation law will always be of the solar-type. The simultaneous solution of the Reynolds equation and the corresponding energy equation provides rotation profiles, meridional circulation patterns, and pole--equator temperature differences that are relatively close to  observations \citep{KR99,KR11}. The results are similar  for all rotation rates and the number of tuning parameters is relatively low.
 
 There are, however, increasing observational indications for the existence of antisolar rotation laws where the equator rotates slower than the polar regions. The results of  observations    \citep{ME16} as well as of   numerical simulations  \citep{GY14,VW18}  suggest that stars with slow rotation possess antisolar rotation laws.
Based on photometric Kepler/K2 data from the open cluster M67,  \cite{BG18} also argue in favor of the appearance of antisolar rotation laws for slow rotators with large Rossby numbers.

 We assume  that because of the often-stated  positivity of the function $H$  the horizontal $\Lambda$ effect  always transports angular momentum towards the equator in favor of an  accelerated equator. If, however,   a rotation law with a negative radial gradient  exists then the rotation-induced  off-diagonal components 
 of the eddy viscosity tensor such as  $\hatnuT$ transport angular momentum towards the poles in favor of a polar vortex (``antisolar''). 
 This also implies that a strictly   radius-dependent rotation law $\Om=\Om(r)$ can never  exist in (slowly) rotating convection zones. After the Taylor-Proudman theorem, the rotation will  tend to produce $z$-independent rotation laws, and therefore a negative radial shear of the angular velocity is  {always} accompanied by slightly accelerated polar regions.
 
We compute the cross-correlations  $\Qrt$, $\Qrf$, and  $\Qtf$ from numerical simulations of convection in a rotating box. The averaging process  concerns the horizontal planes, hence only  radial shear can influence the cross-correlations  (see Eqs. (\ref{dr01}), (\ref{dr02}), and (\ref{correlation3})). For fast rotation, the well-known findings of positive $\Qtf$ and negative  $\Qrt$, $\Qrf$ are reproduced (northern hemisphere). For slow rotation however,   the signs of  both $\Qrt$ and   $\Qtf$ change almost simultaneously, meaning that  the angular momentum is now transported to the poles and the heat is transported to the equator. The resulting warm equator   leads to a counterclockwise meridional circulation which  transports the angular momentum to the poles. The new signs of the quantities  $\Qrt$ and $\Qtf$ may thus lead to antisolar differential rotation.
 
This behavior, however,  is due to   the appearance of  large-scale  flows in zonal and in meridional directions. The zonal  flow $U_z$ mimics differential rotation with negative radial gradient. Via the off-diagonal viscosity in Eq. (\ref{dr02}), for positive $\hatnuT$, a negative contribution to $\Qtf$ leads to overcompensation of the positive but small values of $H$.
A similar effect happens for $\Qrt$ where the negative radial  gradient of $U_y$ combined with the positive eddy viscosity $\nuT$ provides positive contributions to the negative cross-correlation (\ref{Q41}). 
From  the associated  numerical values, with Eq.  (\ref{nuT})   $\nuT$ can also be calculated  leading to $\nuT/\nu\simeq 240$, or to  the  (reasonable) dimensionless quantity $\alpha_{\rm vis}\lsim 0.5$. The simulations lead to the relation $\hatnuT/\nuT\simeq 3 \tau^2_{\rm corr}$ for the ratio of the rotation-induced off-diagonal viscosity term and the standard (diagonal) eddy viscosity.

We also studied the influence of the radial shears of the large-scale flows $U_y$ and $U_z$ on the cross-correlations $Q_{xy}$ and $Q_{yz}$  with numerical experiments where the  $U_y$ and $U_z$ can  artificially be suppressed. In these cases, for all rotation rates, our calculations led to negative $Q_{xy}$ and positive $Q_{yz}$. Without  large-scale flows    the analytical results are confirmed, namely that 
$\Qrt<0$ and $\Qtf>0$ for the northern hemisphere formulated in spherical coordinates. 
For slow rotation, both signs are changed
if the large-scale shear flows are allowed to back-react.

In summary, we show that  a so-far neglected rotation-induced off-diagonal eddy viscosity term  combined with rotation laws with a negative radial gradient (subrotation, as existing in slowly rotating stars) is able to produce  differential rotation of the antisolar type. This result complies with the (numerical) findings of  \cite{VW18}  that negative radial $\Om$-gradients and antisolar differential rotation are  closely related. Therefore, if    new observations confirm the existence of decelerated equators at the surface of  slowly rotating stars  then we shall  better understand the nature of the active eddy viscosity tensor and the underlying turbulence model.

\begin{acknowledgements}
PJK acknowledges the computing resources provided by CSC -- IT Center
for Science, who are administered by the Finnish Ministry of
Education; of Espoo, Finland, and the Gauss Center for Supercomputing
for the Large-Scale computing project ``Cracking the Convective                                                               
Conundrum'' in the Leibniz Supercomputing Centre's SuperMUC
supercomputer in Garching, Germany. This work was supported in part by
the Deutsche Forschungsgemeinschaft Heisenberg programme (grant No.\
KA 4825/1-1; PJK) and the Academy of Finland ReSoLVE Centre of
Excellence (grant No.\ 307411; PJK).
\end{acknowledgements}

\bibliographystyle{aa}
\bibliography{pressure}

\end{document}